\pdfoutput=1 
\documentclass[fleqn,preprint,number,sort&compress,11pt]{elsarticle} 

\usepackage{multirow}
\journal{CMAME}

\usepackage{cmame_config}

\begin{document}
    
    
    \definecolor{MyDarkBlue}{rgb}{1, 0.9, 1}
    \lstset{language=Matlab,
        basicstyle=\footnotesize,
        commentstyle=\itshape,
        stringstyle=\ttfamily,
        showstringspaces=false,
        tabsize=2}
    \lstdefinestyle{commentstyle}{color=\color{green}}
    
    \theoremstyle{remark}
    \newtheorem{thm}{Theorem}[section]
    \newtheorem{rmk}[thm]{Remark}
    
    
    
    \newcommand\restr[2]{{
            \left.\kern-\nulldelimiterspace 
            #1 
            \vphantom{\big|} 
            \right|_{#2} 
    }}
    
    
    

        
\section*{Highlights}
\begin{itemize}
    \item Efficiency of topology optimization algorithm for metamaterials design are improved by coupling with reduced order models.
    \item Isogeometric analysis based on B{\'e}zier  extraction provide solutions with smoother manifold for constructing the reduced basis.
    \item Sensitivity calculation methods including the residual error are derived in the level set method.
\end{itemize}

\begin{frontmatter}
    
    \title{Three-dimensional topology optimization of auxetic metamaterial using isogeometric analysis and model order reduction}
    
    \author[hannover]{Chuong Nguyen}
    \author[hannover]{Xiaoying Zhuang\corref{cor2}\fnref{XiaoyingZhuang}}
    \author[cachan]{Ludovic Chamoin\corref{cor2}\fnref{LudovicChamoin}}
    \author[hutechvietnam]{Hung Nguyen-Xuan}
    \author[tongji]{Xianzhong Zhao}
    \author[weimar]{Timon Rabczuk\corref{cor2}\fnref{TR}}
    
    \cortext[cor2]{Corresponding authors}
    \address[hannover]{Institute of Continuum Mechanics, Leibniz Universit\"{a}t Hannover, Hannover, Germany }
    \address[cachan]{LMT, ENS Paris-Saclay, CNRS, Universit\'{e} Paris-Saclay 61 Avenue du Pr\'{e}sident Wilson, 94235 Cachan, France}
    \address[hutechvietnam]{CIRTech Institute, Ho Chi Minh City University of Technology (HUTECH), Ho Chi Minh City, Vietnam}
    \address[tongji]{College of Civil Engineering, Tongji University, China}
    \address[weimar]{Institute of Structural Mechanics, Bauhaus-Universit\"{a}t Weimar, Weimar, Germany}
    
    \fntext[XiaoyingZhuang]{\url zhuang@ikm.uni-hannover.de}
    \fntext[LudovicChamoin]{\url ludovic.chamoin@ens-paris-saclay.fr}
    \fntext[TR]{\url timon.rabczuk@uni-weimar.de}
    
    \begin{abstract}
        In this work, we present an efficiently computational approach for designing material micro-structures  by means of topology optimization. The central idea relies on using the isogeometric analysis integrated with the  parameterized level set function  for numerical homogenization, sensitivity calculation and optimization of the effective elastic properties. Design variables, which are level set values associated with control points, are updated from the optimizer and represent the geometry of the  unit cell. We further improve the computational efficiency in each iteration by employing reduced order modeling when solving  linear systems of the equilibrium equations. We construct a reduced basis by reusing computed solutions from previous optimization steps, and a much smaller linear system of equations is solved on the reduced basis. Two- and three-dimensional numerical results show the effectiveness of the topology optimization algorithm coupled with the reduced basis approach in designing metamaterials.
    \end{abstract}
    
    \begin{keyword} 
        metamaterials, topology optimization, reduced order model, isogeometric analysis,  B{\'e}zier extraction
    \end{keyword}
    
\end{frontmatter}


\section{Introduction}
Metamaterials are classified as  artificial composites possessing extraordinary mechanical behaviors that are not available in  nature. They have received attention in scientific applications and engineering over the last  decades. Typically the class of auxetic metamaterials having properties of  negative Poisson's ratio (NPR)  have shown potential applications from biomedical to defense problems \cite{Lake1993,Evans2000}. Modifying the effective Poisson's ratio, which affects kinematic deformations, can enhance the performance of embedded structures. The material functionalities are attained by rationally designing their microstructure layouts. Among material design methods, shape and topology optimization  for primitive cells has been considered as a prominent method for materials functionality design. 

Topology optimization generally involves iterative searching for material distribution within  design domains. In the density-based topology optimization,  the void and solid regions of material are represented by density values, which are also considered as design variables. Among the density approaches, the evolutionary structural optimization (ESO) \cite{Xie1993,Xie1997} and solid isotropic material with penalization (SIMP) method \cite{Bendsoe1999} are frequently used in structural and material optimization \cite{Sigmund1994b,Larsen1997design} due to the efficiency and easy implementation. Alternatively, the level set method represents structural boundaries by a zero level set function, which enables flexible changes in shape and topology with  distinct material interfaces. Various optimization problems have been successfully implemented with the level set approach, e.g., structure compliance minimization \cite{Allaire2002,Allaire2016}, frequency response problem \cite{Allaire2017}, or design of metamaterials with negative Poisson's ratio \cite{Vogiatzis201715,Wang2004469}.

In microstructure topology optimization, the effective properties of unit cells are determined by a numerical homogenization procedure. This involves solving linear systems of equilibrium equations and requires high computational cost. In order to circumvent this issue, reduced order modeling using Krylov subspaces  \cite{Amir2011,Wang2007a} has been proposed to enhance the efficiency. Large linear systems are solved by an iterative solver and using a search space from the previous linear system. The approach assumes that numerical stiffness from two consecutive iterations are slightly different due to  minor changes in shapes or topology of structures. Alternatively,  the reduced-basis approaches approximate high-dimensional solutions by the projection of the original model into a lower-dimensional subspace which is spanned by global basis functions or truncated modes. This surrogate model was employed in nonlinear structural optimization \cite{Xia2014}, multiscale homogenization \cite{Hernandez2014}. Similarly, Gogu \cite{Gogu2015} proposed a reduced basis constructed from displacements which are calculated from a set of similar configurations. The original linear system is then projected on the reduced basis with a smaller size. The method has shown much improvement in structural topology optimization.  

The motivation for using isogemetric analysis \cite{Hughes2005,Cottrell2009isogeometrics}  so far is due to higher-order approximation with  the B-Spline or NURBS basis functions which are compatible with CAD tools. It enables to model the geometry in computer-aided design  exactly, and the basis functions used to represent the geometry can directly approximate solution fields in numerical computation. Recently, this framework  has also been used in various optimization problems, i.e., \cite{Wang2016b} with level set functions,  \cite{Costa2019} using the SIMP approach in structural optimization and \cite{Gao2019b} for auxetic metamaterials. Several works in \cite{Manzoni2015,Zhu2017}   introduced IGA coupled with reduced order models (ROM) to alleviate the computational cost for parameterized geometry model, and have shown an improvement in both accuracy and efficiency compared to the finite element method. We also mention the shape optimization problems in which the  geometry parameters assigned as design variables are efficiently optimized with the combination of reduced order approaches. For example in  \cite{Ammar2013a}, the proper generalized decomposition (PGD) was introduced to compute in an offline phase the parametric solutions for selected parameters which involve the sensitivity information provided. A similar strategy for parameter optimization may be applied with the isogeometric framework, such as   \cite{Chamoin2019a}, where model reduction based on PGD provides  parametric solutions with much lower computation cost. 

In this work, we exploit advantages of  the parameterized level set method using the isogeometric setting for metamaterials design. This approach possibly allows using basis functions with higher orders to approximate the displacement field and the level set function. We further integrate reduced order modeling to improve the efficiency of the topology optimization algorithm. The proper coupling of reduced order techniques  in an isogeometric framework is attractive and feasible due to the fact that the reduced basis is constructed from a smoother manifold of the isogeometric solution, and the optimized geometry is subsequently generated   from   the reduced solutions with saving computational time. We provide sensitivity analysis involving the reduced solutions in the level set framework. Metamaterial structures are designed in  two and three dimensions to show the efficiency and reliability of the approach.

The paper is organized as follows: in \cref{sect_homogenization_method}, we briefly discuss numerical homogenization methods and formulation for the effective elasticity tensor. \cref{sect_level_set_function} introduces the parameterized version of the level set function. In \cref{sect_IGA_discretization}, the discretized formulation with the isogeometric analysis using B{\'e}zier extraction  for the numerical analysis  is given.  Afterward,  the construction of the reduced basis is reviewed in \cref{sect_Reduced_order_modelling},  and  in \cref{sect_sensitivity_analysis_reduced_model} the sensitivity analysis involving reduced solutions of effective coefficients is described.  Numerical examples are presented in \cref{sect_numerical_examples} and followed by the conclusion in \cref{sect_conclusion}. 


\section{Homogenization method}\label{sect_homogenization_method}
Heterogeneity of composite materials by nature requires  fine scale discretization  to capture the geometric details; this leads to large finite element models and increased computational burden. An alternative material modeling is to replace heterogeneity with an effective homogeneous model by considering a representative volume that can describe the equivalent properties at the macroscopic scale. The classical approach is  based on the assumption of periodic arrangement of the micro-structures, and it also considers that the length scale of the periodic structures is small when compared to dimensions of the macroscopic structure. The study of macroscopic properties is conventionally replaced by considering  of a single unit cell alone.

Several analytical approaches are available to evaluate the effective properties of composites. For example, considering the volume fraction of a single inclusion embedded into an infinite matrix material, the effective properties were  derived  by Eshelby \cite{Eshelby1957}. Further developments from this approach and widely employed in modeling homogenized material can be found in the work by Mori and Tanaka \cite{Mori1973a}, or self-consistent scheme by Hill \cite{Hill1965}. These analysis methods mainly deal with simple geometries of inclusions, i.e., circles or ellipses. In the context of topology optimization, complex geometries of the unit cell frequently occur and the analytical models are not able to predict effective properties of composites. 

Numerical approaches have been developed to address the problem. From macroscopic strains, stresses and strain energy density, the effective elasticity of inhomogeneous materials was derived using a direct average method \cite{Becker2001,Yang2004comparative,Sanchez-Palencia1986}. Similarly, Zhang et al. \cite{Zhang2007}  used the strain energy-based methods to predict effective properties in microstructure topology optimization. The work of Guedes and Kikuchi \cite{Guedes1990}, or Sanchez-Palencia \cite{Sanchez-Palencia1986} introduced a rigorous mathematical theory for deriving effective elasticity coefficients by using weak-form of equilibrium equation and limits theory, namely asymptotic homogenization method. Another mathematical derivation for homogenized elasticity which is based on the governing equations with strong formulation can be found in the work by Zhuang et al. \cite{Zhuang2015536}, and a similar framework for multi-field problems was recently investigated by Fantoni et al. \cite{Fantoni2017}.

The design of metamaterials in this work relies on the asymptotic homogenization method. Effective properties across a unit cell are obtained from a homogenization procedure with assumptions of periodicity and scale separation over the unit cell with overall macroscopic dimensions. A brief explanation of the asymptotic homogenization method used in this paper is given in the following and details of derivation can be found in \cite{Guedes1990,Hassani2012homogenization}. 

Due to local changes of material properties, the displacement expression is dependent on the small parameter $\epsilon$  and written as an asymptotic expansion: 
\begin{equation}\label{equ_asymptotic_expansion_of_displacement}
{\renewcommand{\arraystretch}{1.0}
    \begin{array}{*3{>{\displaystyle}l}} 
    \mathbf{u}^{\epsilon} (\mathbf{x},\mathbf{y},\epsilon) = \mathbf{u}^{0}(\mathbf{x}) + \epsilon \mathbf{u}^{1}(\mathbf{x},\mathbf{y}) + \epsilon^{2} \mathbf{u}^{2}(\mathbf{x},\mathbf{y})+ ... & \text{with} \quad \mathbf{y}=\mathbf{x}/\epsilon,
    \end{array}
}
\end{equation}
$\mathbf{x}$ and $\mathbf{y}$ are macroscopic and microscopic (unit cell) coordinates respectively. The displacement $\mathbf{u}^{\epsilon}$ of the composite body is solution to the variational problem
\begin{equation}\label{equ_weak_form_macroscopic_equation}
{\renewcommand{\arraystretch}{1.0}
    \begin{array}{*3{>{\displaystyle}l}} 
    \int\limits_{\Omega} \pmb{\varepsilon}(\mathbf{u}^{\epsilon}):\mathbb{C}:\pmb{\varepsilon}(\mathbf{v})\  \, d\Omega  = \int\limits_{\Omega} \mathbf{b}\cdot \mathbf{v} \, d\Omega +  \int\limits_{\Gamma} \tilde{\mathbf{t}}\cdot \mathbf{v}  \, d\Gamma , 
    \end{array}
}
\end{equation}
where $\mathbf{v}$ is the test function and $\pmb{\varepsilon}$ is the mechanical strain.  Body forces $\mathbf{b}$ and surface tractions $\tilde{\mathbf{t}}$ are assumed not to vary over the microscopic domain.  In the following procedure only the first and second variation terms of the displacement expansion  \eqref{equ_asymptotic_expansion_of_displacement} are employed to derive the effective elasticity tensor, and due to the linearity of the problem the second term can be written as
\begin{equation}\label{}
{\renewcommand{\arraystretch}{1.0}
    \begin{array}{*3{>{\displaystyle}l}} 
    \mathbf{u}^{1}\left(\mathbf{x},\mathbf{y}\right) = \pmb{\chi}\left(\mathbf{x},\mathbf{y}\right) \pmb{\varepsilon}(\mathbf{u}^{0}\left(\mathbf{x}\right)),
    \end{array}
}
\end{equation}
where $\pmb{\chi}\left(\mathbf{x},\mathbf{y}\right)$ is the characteristic displacement. By substituting the asymptotic expansion \eqref{equ_asymptotic_expansion_of_displacement}  into \eqref{equ_weak_form_macroscopic_equation} and applying the asymptotic analysis of the periodic functions in a unit cell  \cite{Guedes1990},  the following equilibrium equation is obtained
\begin{equation}\label{equ_equilibrium_equation_in_unit_cell}
{\renewcommand{\arraystretch}{1.0}
    \begin{array}{*3{>{\displaystyle}l}} 
    \int\limits_{Y}  \varepsilon_{pq}\left(\pmb{\chi}^{mn}\right) C_{pqrs}(\mathbf{y}) \, \varepsilon_{rs}(\mathbf{v})  \, dY =\int\limits_{Y} \varepsilon^{0,mn}_{pq}  C_{pqrs}(\mathbf{y}) \, \varepsilon_{rs}(\mathbf{v})  \, dY & \forall \mathbf{v} \in V_{0} \subset H^{1},
    \end{array}
}
\end{equation}
where $\varepsilon^{0,mn}$ is the unit strain. The superior index $mn$ indicates the test cases. For a two-dimensional problem, there are three independent unit strains  $\pmb{\varepsilon}^{0,11} = [1,0,0]$, $\pmb{\varepsilon}^{0,22} = [0,1,0]$ and $\pmb{\varepsilon}^{0,12} = [0,0,1]$. The characteristic displacements $\pmb{\chi}^{mn}$ are obtained numerically under periodic condition in the unit cell $Y$. Having these local solutions, the effective elasticity tensor $\mathbb{C}^{H}$ of the unit cell is calculated by 
\begin{equation}\label{equ_effective_elastic_tensor}
{\renewcommand{\arraystretch}{1.0}
    \begin{array}{*3{>{\displaystyle}l}} 
    C^{H}_{ijkl}=\frac{1}{|Y|}\int\limits_{Y} (\varepsilon^{0,ij}_{pq} - \varepsilon_{pq}(\pmb{\chi}^{ij})) C_{pqrs}(\mathbf{y}) (\varepsilon^{0,kl}_{rs} + \varepsilon_{rs}(\pmb{\chi}^{kl}))  \, d\Omega,
    \end{array}
}
\end{equation}
which essentially corresponds to the average of strain energy in the unit cell.

\section{Level set function}\label{sect_level_set_function}
\begin{figure}[ht!]
    \centering 
    \setlength\figureheight{5.0cm}
    \setlength\figurewidth{6.0cm}
    \setlength\tabcolsep{0.0pt} 
    \begin{tabular}{ccc}
        \includegraphics[width=0.225\linewidth]{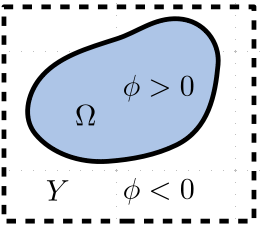} & \hspace{2cm} &\includegraphics[width=0.225\linewidth]{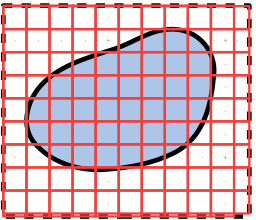}\\
        (a) & & (b)
    \end{tabular}
    \caption{Unit cell geometry represented by a level set function (a) and fixed mesh on a square domain (b).}
    \label{fig_Level_set_function_and_fixed_square_domain}
\end{figure}

In this section we briefly discuss the fundamentals of the level set method in structural optimization. The material distribution in the unit cell is represented via an implicit level set function defined in a reference domain $Y$  as shown in \autoref{fig_Level_set_function_and_fixed_square_domain}(a) with 
\begin{equation}\label{}
{\renewcommand{\arraystretch}{1.0}
    \begin{array}{*3{>{\displaystyle}l}} 
    \left\{\begin{array}{*3{>{\displaystyle}l}} 
    \phi(\mathbf{x}) >0, \quad \mathbf{x}\in \Omega \, \text{(solid)}\\
    \phi(\mathbf{x}) <0, \quad \mathbf{x} \in Y/\Omega \, \text{(void)}\\
    \phi(\mathbf{x}) =0, \quad \mathbf{x}\in \partial\Omega \, \text{(boundary)}.
    \end{array}  
    \right.      
    \end{array}
}
\end{equation}
The interfaces between void and solid parts are considered as design boundaries. The evolution of the design boundaries in the optimization process are handled by dynamic moving of the zero level set $\phi(\mathbf{x}(\tau),\tau)=0$.  The dynamic change in time is governed by the Hamilton–Jacobi (H-J) equation
\begin{equation}\label{}
{\renewcommand{\arraystretch}{1.0}
    \begin{array}{*3{>{\displaystyle}l}} 
    \frac{\partial \phi}{\partial t} - V_{n} |\nabla\phi| = 0, \quad V_{n} = \frac{d\mathbf{x}}{d t} \cdot \frac{\nabla\phi}{|\nabla\phi|},
    \end{array}
}
\end{equation}
where $V_n$ is the normal velocity component. In the steepest descend method,  $V_{n}$ is chosen such that the gradient of the objective function is negative to ensure the objective function is minimized. Conventionally, the level set function is updated by solving the Hamilton–Jacobi equation with a finite difference scheme \cite{Wang2003227,Allaire2004363}, and consequently a new design boundary is obtained and used to proceed with a next optimization iteration. It is noted that a restriction of step sizes is necessary to ensure numerical stability when solving the H-J equation with finite difference schemes, i.e., Courant–Friedrichs–Lewy (CFL) condition. Furthermore, the signed distance characteristic of the level set function should be maintained during the optimization procedure to avoid the flatness or steepness; therefore, additional re-initialization steps are provided periodically after several iterations \cite{Osher2003,Allaire2004363}.

In this present work, the parameterized level set method (PLSM) \cite{Belytschko2003,Wang2004469} is used. The level set function is approximated by
\begin{equation}\label{equ_parameterized_lsf}
{\renewcommand{\arraystretch}{1.0}
    \begin{array}{*3{>{\displaystyle}l}} 
    \phi(\mathbf{x},\tau) = \sum\limits_{I}^{ncp} R_{I}(\mathbf{x})\,\alpha_{I}(\tau) =  \mathbf{R}(\mathbf{x}) \cdot \pmb{\alpha}(\tau),
    \end{array}
}
\end{equation}
and the H-J PDE equation reduces to an ordinary differential equation
\begin{equation}\label{equ_H_J_ODE}
{\renewcommand{\arraystretch}{1.0}
    \begin{array}{*3{>{\displaystyle}l}} 
    \mathbf{R}^{T} \frac{d\pmb{\alpha}}{d\tau} - V_{n} |(\nabla \mathbf{R})^T \pmb{\alpha}| =0 & \text{or} \quad V_{n}=\frac{\mathbf{R}^{T} \dot{\pmb{\alpha}}}{|(\nabla \mathbf{R})^T \pmb{\alpha}|}
    \end{array}
}
\end{equation}
where $R_{I}(\mathbf{x})$ are B-spline basis functions in the isogeometric concept. These functions will be discussed in \cref{sect_IGA_discretization}. Herein, the expansion coefficients $\alpha_{I}$ are considered as design variables. The level set function is updated by changing values of the coefficients $\alpha_{I}$ during the optimization procedure. A gradient-based mathematical programming optimizer is used to optimize the design variables. Solving the H-J equation is avoided and the limit of step sizes to ensure numerical stability is unnecessary, so that the efficiency is improved. Furthermore, there is no re-initialization step to keep the level set function to be a signed distance function.

For a unit cell with two material phases, solid and void, as shown in \cref{fig_Level_set_function_and_fixed_square_domain}(a), we can express spatial dependence in  the elasticity tensor with a Heaviside function $\mathcal{H}(\phi)$ of the level set value. Equations \eqref{equ_equilibrium_equation_in_unit_cell} and \eqref{equ_effective_elastic_tensor} are written as 
\begin{equation}\label{equ_equilibrium_equation_in_unit_cell_with_lsf}
{\renewcommand{\arraystretch}{1.0}
    \begin{array}{*3{>{\displaystyle}l}} 
    \int\limits_{Y} \varepsilon_{pq}(\pmb{\chi}^{ij}) C_{pqrs}\mathcal{H}(\phi(\mathbf{y})) \, \varepsilon_{rs}(\mathbf{v})  \, d\Omega =\int\limits_{Y} \varepsilon^{0,ij}_{pq}  C_{pqrs}\mathcal{H}(\phi(\mathbf{y})) \, \varepsilon_{rs}(\mathbf{v})  \, d\Omega ,
    \end{array}
}
\end{equation}

\begin{equation}\label{equ_effective_elastic_tensor_with_lsf}
{\renewcommand{\arraystretch}{1.0}
    \begin{array}{*3{>{\displaystyle}l}} 
    C^{H}_{ijkl}=\frac{1}{|Y|}\int\limits_{Y} \left(\varepsilon^{0,ij}_{pq} - \varepsilon_{pq}(\pmb{\chi}^{ij})\right) C_{pqrs}\mathcal{H}(\phi(\mathbf{y})) \left(\varepsilon^{0,kl}_{rs} - \varepsilon_{rs}(\pmb{\chi}^{kl})\right)  \, d\Omega,
    \end{array}
}
\end{equation}
and for the volume of solid part
\begin{equation}\label{}
{\renewcommand{\arraystretch}{1.0}
    \begin{array}{*3{>{\displaystyle}l}} 
    V =  \int\limits_{Y} \mathcal{H}(\phi(\mathbf{y})) d\Omega.
    \end{array}
}
\end{equation}
To obtain the derivative of the Heaviside function numerically, we use a regularized Heaviside function of the form  
\begin{equation}\label{}
{\renewcommand{\arraystretch}{1.0}
    \mathcal{H}(\phi)= \left\{\begin{array}{*3{>{\displaystyle}l}} 
    \rho_{\min} & \phi < \xi\\
    \frac{3}{4}(\frac{\phi}{\xi}-\frac{\phi^3}{3\xi^3}) + \frac{1}{2} & -\xi \leq \phi \leq \xi\\
    1 & \phi > \xi
    \end{array}\right.
}
\end{equation}
where $\xi$ is the smooth length, and $\rho_{\min}=10^{-6}$ is chosen to avoid singularity in the numerical stiffness. 


\section{Isogeometric analysis with B{\'e}zier extraction}\label{sect_IGA_discretization}
This section gives a brief description of the isogeometric analysis (IGA) concept \cite{Cottrell2009isogeometrics} and the Bernstein-B{\'e}zier representation   \cite{Borden2010bezierextraction} to non-uniform rational basis functions (NURBS). Similar to the finite element approach, the actual geometry in the physical space is represented in a parameter space by using a mapping  
\begin{equation}\label{equ_finite_basis_functions}
{\renewcommand{\arraystretch}{1.0}
    \begin{array}{*3{>{\displaystyle}l}} 
    \mathbf{T}(\pmb{\xi}) = \sum\limits_{I=1}^{ncp} R_{I}(\pmb{\xi}) \mathbf{P}_{I},
    \end{array}
}
\end{equation}
with a mesh of $ncp $ control points $\mathbf{P}_{I}$ and associated weight $w_{I}$. The NURBS basis functions $R_{I}(\pmb{\xi})$ are defined as
\begin{equation}\label{}
{\renewcommand{\arraystretch}{1.0}
    \begin{array}{*3{>{\displaystyle}l}} 
    R_{I}(\pmb{\xi})  = \frac{w_{I} \, N_{I}(\pmb{\xi})}{W} = \frac{w_{I} \, N_{I}(\pmb{\xi})}{\sum_{J=1}^{N} N_{J}(\pmb{\xi}) \, w_{J}} ,
    \end{array}
}
\end{equation}
where the B-spline functions $N_{I}(\pmb{\xi})$ of degree $p$ are defined by the knot-vectors on a parameter domain $\Omega_{\xi} = [0,1]^d$:
\begin{equation}\label{}
{\renewcommand{\arraystretch}{1.0}
    \begin{array}{*3{>{\displaystyle}l}} 
    \Xi^{i}:=\{\xi^{i}_{1},...,\xi^{i}_{n_{i}+p+1}\}, & 0=\xi^{i}_{1}\leq \xi^{i}_{2}\leq \xi^{i}_{n_{i}+p+1}= 1 & \text{and}\quad i=1,...,d,
    \end{array}
}
\end{equation}
and $n_{i}$ is the number of B-spline functions in each spatial direction. 

\begin{figure}[ht!]
    \centering 
    \setlength\figureheight{5.0cm}
    \setlength\figurewidth{6.0cm}
    \setlength\tabcolsep{0.0pt} 
    \begin{tabular}{cc}
        \includegraphics[width=0.5\linewidth]{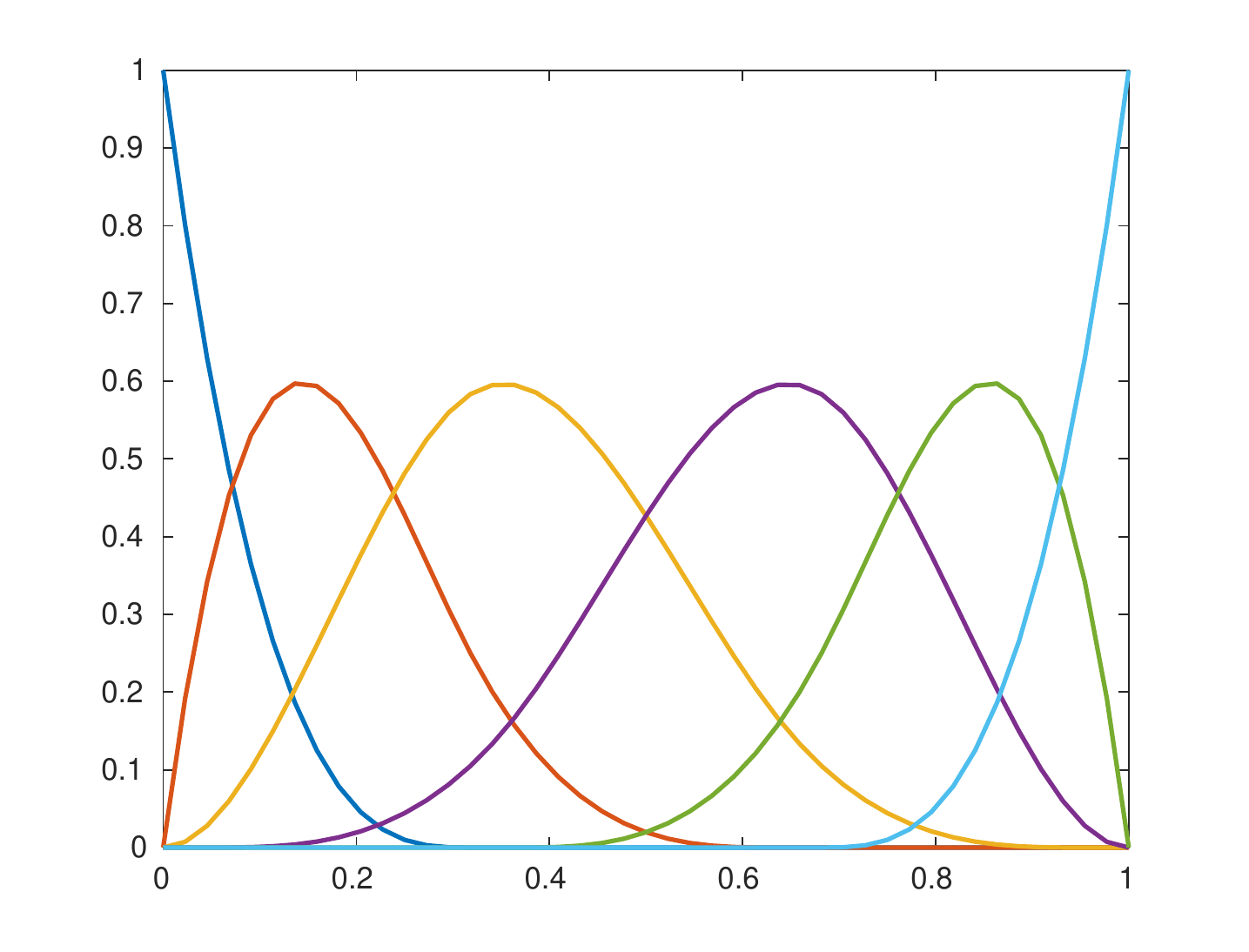} &  \includegraphics[width=0.5\linewidth]{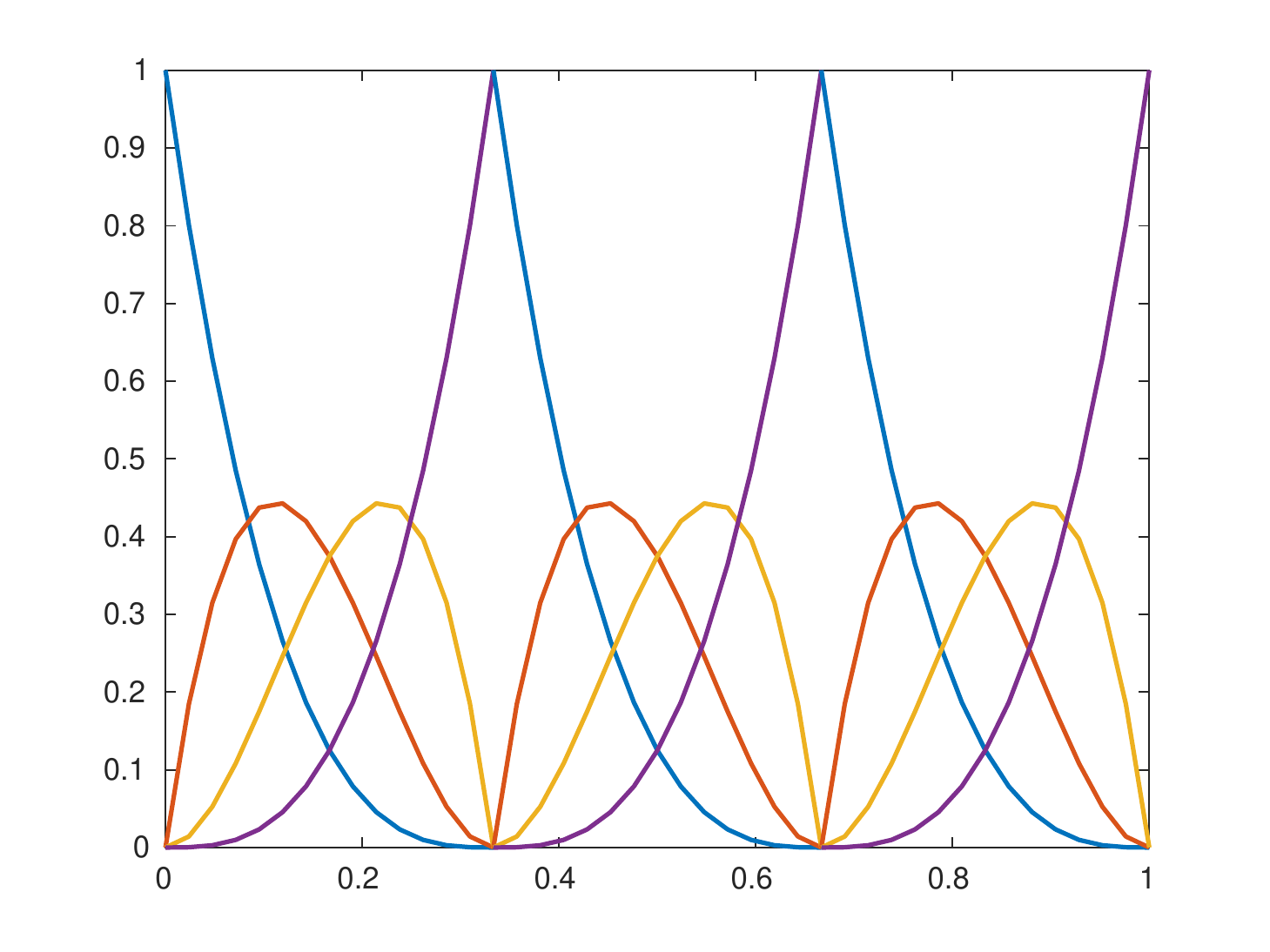}
    \end{tabular}
    \caption{B-spline basis functions (left) and their corresponding Bernstein-B{\'e}zier representation (right).}
    \label{fig_bezier_bspline_function}
\end{figure}

B-spline basis functions in the parameter space span in several elements (knot spans), which makes numerical implementation more difficult. In order to use element-wise structures similar to the standard finite element method, B-spline basis functions are rewritten as  a linear combination of the Bernstein polynomials $\mathbf{B}(\pmb{\xi})$ defined on a parent element domain $[-1,1]^d$, generating B{\'e}zier elements which have  $C^0$-continuity (see \cref{fig_bezier_bspline_function})
\begin{equation}\label{}
{\renewcommand{\arraystretch}{1.0}
    \begin{array}{*3{>{\displaystyle}l}} 
    \mathbf{N}^{e}=\mathbf{C}^{e} \mathbf{B}(\pmb{\xi}).
    \end{array}
}
\end{equation}
$\mathbf{N}^{e}$ is a set of $nb$ B-spline basis functions associated to the B{\'e}zier element, i.e., $\mathbf{N}^{e}=[N^{e}_{1},...,N^{e}_{nb}]$. The localized extraction operator $\mathbf{C}^{e}$ is constructed from the knot vector  and is independent of control points as well as the B-spline basis functions. Similarly, NURBS basis functions within a B{\'e}zier element have the form:
\begin{equation}\label{}
{\renewcommand{\arraystretch}{1.0}
    \begin{array}{*3{>{\displaystyle}l}} 
    \mathbf{R}^{e}(\pmb{\xi}) = \frac{ \mathbf{W}^{e}\mathbf{C}^{e}\mathbf{B}(\pmb{\xi})}{\sum_{J=1}^{nb} w_{J} N^{e}_{J}(\pmb{\xi})} = \frac{ \mathbf{W}^{e}\mathbf{N}^{e}(\pmb{\xi})}{W(\pmb{\xi})}, 
    \end{array}
}
\end{equation}
where $\mathbf{W}^{e}$ is a diagonal matrix of weights. The implementation of IGA using the B{\'e}zier extraction allows the possibility of using existing routines in the finite element codes, i.e, assembly and post-processing. 

For a unit cell discretized by $nel$ B{\'e}zier elements as shown in \cref{fig_Level_set_function_and_fixed_square_domain}(b), the equilibrium equation \eqref{equ_equilibrium_equation_in_unit_cell} has the following expression in matrix form
\begin{equation}\label{equ_equilibrium_equation_for_test_case}
{\renewcommand{\arraystretch}{1.0}
    \begin{array}{*3{>{\displaystyle}l}} 
    \mathbf{K} \, \pmb{\chi}^{mn}=\mathbf{F}^{mn},
    \end{array}
}
\end{equation}
where the global stiffness matrix $\mathbf{K}$ and force vector $\mathbf{F}^{mn}$ are assembled from corresponding element matrix and vector
\begin{equation}\label{}
{\renewcommand{\arraystretch}{1.0}
    \begin{array}{*3{>{\displaystyle}l}} 
    \mathbf{k}^{e}=\int\limits_{\Omega^{e}} \mathbf{B}^{T} \mathbf{C} \, \mathbf{B} \, d\Omega^{e}, & \mathbf{f}^{e,mn}=\int\limits_{\Omega^{e}} \mathbf{B}^{T} \mathbf{C} \, \pmb{\varepsilon}^{0,mn} \, d\Omega^{e}.
    \end{array}
}
\end{equation} 
Herein, $\mathbf{B}$ is the standard matrix of elastic strain operator in solid mechanics \cite{Zienkiewicz2000thebasis}, and $\mathbf{C}$ is the matrix of elasticity coefficient. The effective elasticity tensor is calculated from characteristic displacements:
\begin{equation}\label{}
{\renewcommand{\arraystretch}{1.0}
    \begin{array}{*3{>{\displaystyle}l}} 
    C^{H}_{ijkl}= \frac{1}{|Y|} \sum\limits_{e}^{nel} \int\limits_{\Omega^{e}} \left[\pmb{\varepsilon}^{0,ij} - \mathbf{B} \pmb{\chi}^{ij}\right]^{T}   \mathbf{C} \, \left[\pmb{\varepsilon}^{0,kl} - \mathbf{B} \pmb{\chi}^{kl}\right].
    \end{array}
}
\end{equation}

\section{Reduced order modelling}\label{sect_Reduced_order_modelling}
At each optimization iteration, the characteristic displacements $\pmb{\chi}$ are obtained by solving the system of equations \eqref{equ_equilibrium_equation_for_test_case} and the computational cost involved in the inversion of stiffness matrix $\mathbf{K}$ is expensive in large scale problems. We aim at reducing the finite dimensional space $V^{h}_{0}  \subset H^{1} $  by projecting the balance equation to subspace $\tilde{V}^{h}_{0}\subset V^{h}_{0}$ with lower dimension 
\begin{equation}\label{}
{\renewcommand{\arraystretch}{1.0}
    \begin{array}{*3{>{\displaystyle}l}} 
    \dim (\tilde{V}^{h}_{0}) = nb \ll \dim(V^{h}_{0}).
    \end{array}
}
\end{equation}
The task is to construct global ansatz functions $\theta_{m}(\mathbf{y})$  such that the subspace $\tilde{V}^{h}_{0} = \text{span}\left\{\theta_{1},...,\theta_{nb}\right\}$, and solution fields in the reduced space are defined by
\begin{equation}\label{}
{\renewcommand{\arraystretch}{1.0}
    \begin{array}{*3{>{\displaystyle}l}} 
    \chi^{h}(\mathbf{y}) \simeq \tilde{\chi}^{h}(\mathbf{y})=\sum\limits_{m}^{nb}\varphi_{m}\theta_{m}(\mathbf{y}).
    \end{array}
}
\end{equation}
The functions $\theta_{m}$ define the linear space $\tilde{V}^{h}_{0}$ and are associated to the matrix of coefficients $\pmb{\Phi} \in \mathbb{R}^{N\times nb}$ by
\begin{equation}\label{equ_global_ansatz_function}
{\renewcommand{\arraystretch}{1.0}
    \begin{array}{*3{>{\displaystyle}l}} 
    \theta_{m}(\mathbf{y}) = \sum\limits_{n}^{N} \Phi_{nm} R_{n}(\mathbf{y}),
    \end{array}
}
\end{equation}
where $ R_{n}$ are basis functions defined in \eqref{equ_finite_basis_functions}.  The balance equations \eqref{equ_equilibrium_equation_for_test_case} have the size of $nb \ll N$ in the subspace and are formed as
\begin{equation}\label{equ_linear_equation_on_subspace}
{\renewcommand{\arraystretch}{1.0}
    \begin{array}{*3{>{\displaystyle}l}} 
    \tilde{\mathbf{K}} \tilde{\pmb{\chi}} = \tilde{\mathbf{F}},
    \end{array}
}
\end{equation}
where $\pmb{\chi}=\pmb{\Phi} \tilde{\pmb{\chi}}$, $\tilde{\mathbf{K}} =\pmb{\Phi}^{T}\mathbf{K}\pmb{\Phi}$ and $ \tilde{\mathbf{F}} = \pmb{\Phi}^{T}\mathbf{F}$.

\begin{figure}[ht!]
    \centering 
    \setlength\figureheight{5.0cm}
    \setlength\figurewidth{6.0cm}
    \setlength\tabcolsep{0.0pt} 
    \begin{tabular}{cc}
        \includegraphics[width=1.0\linewidth]{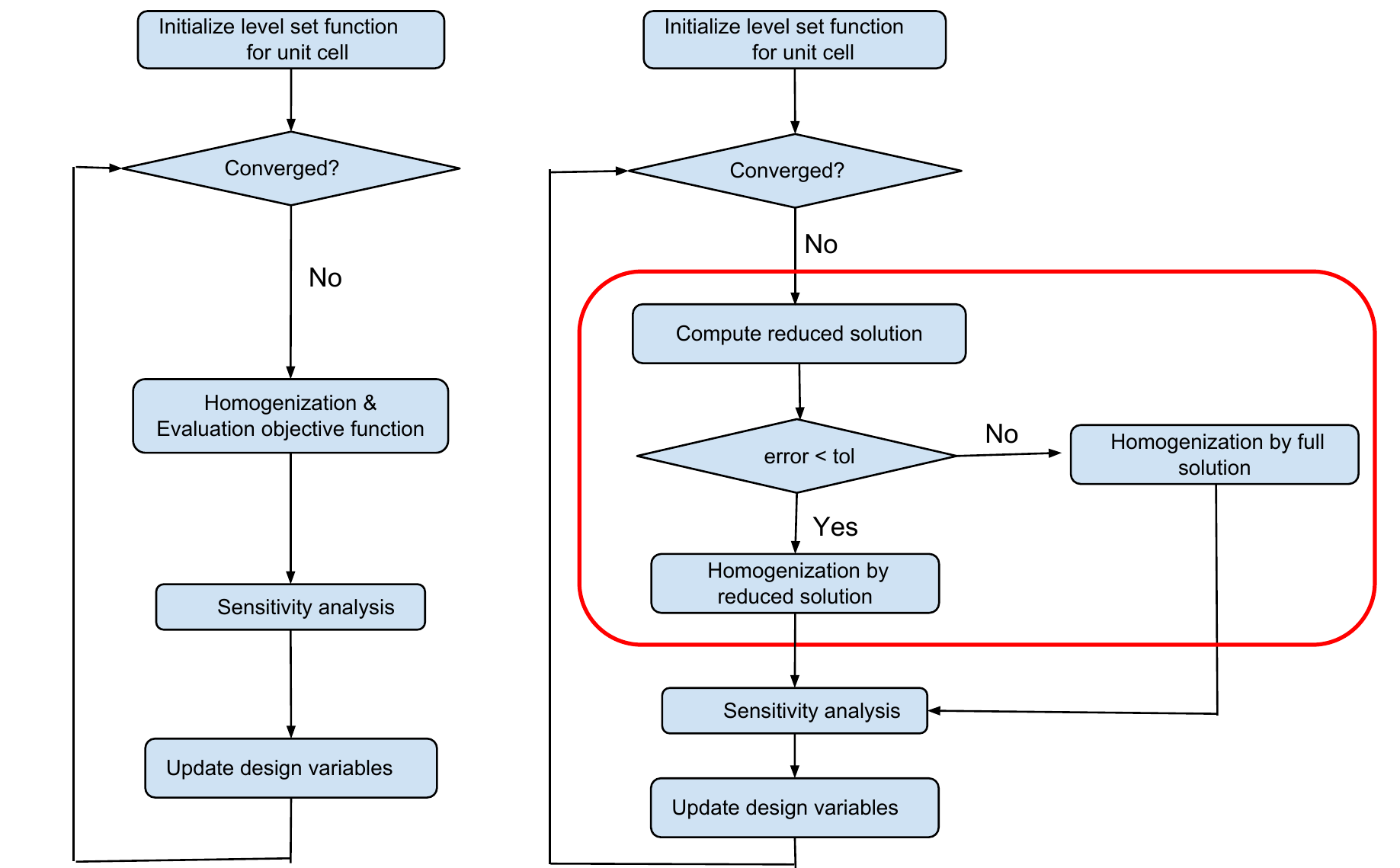}
    \end{tabular}
    \caption{Microstructure topology optimization algorithm with and without reduced basis model.}
    \label{fig_optimization_algorithm_with_reduced_basis}
\end{figure}

The construction of the reduced basis follows the work by Gogu \cite{Gogu2015}. The key idea is to predict a new displacement solution and consequently a new topology by a set of solutions obtained from the previous iterations. We initialize the matrix $\pmb{\Phi}$ with  vectors of coefficients  $\pmb{\phi}_{m}$ calculated from the corresponding solutions $\pmb{\chi}_{m}$ (i.e., displacements associated to control points) in the first $nb$ iterations. The first basis vector is given as
\begin{equation}\label{equ_contructed_RM01}
{\renewcommand{\arraystretch}{1.0}
    \begin{array}{*3{>{\displaystyle}l}} 
    \pmb{\phi}_{1}=\frac{\pmb{\chi}_{1}}{\|\pmb{\chi}_{1}\|}.
    \end{array}
}
\end{equation}
The next basis vector is computed by the Gram-Schmidt orthogonalization and normalization:
\begin{equation}\label{equ_contructed_RM02}
{\renewcommand{\arraystretch}{1.0}
    \begin{array}{*3{>{\displaystyle}l}} 
    \tilde{\pmb{\phi}}_{m+1}= \pmb{\chi}_{m+1} - \sum\limits_{j}^{m}\langle\pmb{\chi}_{m+1},\pmb{\phi}_{j}\rangle \, \pmb{\phi}_{j}\\\\
    \pmb{\phi}_{m+1}=\frac{\tilde{\pmb{\phi}}_{m+1}}{\|\tilde{\pmb{\phi}}_{m+1}\|}, \quad  m=1,...,(nb-1).
    \end{array}
}
\end{equation}
Once the reduced basis is formed, a solution in the next iteration is obtained in the subspace of dimension $nb$ and then transformed back to full space by using \eqref{equ_linear_equation_on_subspace}. In order to assess the accuracy of the solution obtained from the reduced basis, the following error measurement is suggested (see also \cite{Gogu2015})
\begin{equation}\label{equ_error_reduce_solution}
{\renewcommand{\arraystretch}{1.0}
    \begin{array}{*3{>{\displaystyle}l}} 
    error=\frac{\|\mathbf{K}\pmb{\Phi}\tilde{\pmb{\chi}} - \mathbf{F}\|}{\|\mathbf{F}\|} < tol.
    \end{array}
}
\end{equation}
If the error is acceptable, the optimization algorithm is conducted using the reduced solution. Otherwise, a full solution is calculated and used for the computations at the subsequent steps. After each successful iteration, the reduced basis is enriched by adding the new accepted solution. The comparison of conventional optimization algorithms and those with  reduced basis model is shown in \cref{fig_optimization_algorithm_with_reduced_basis}. The two flowcharts have similar procedures except for the decision of using reduced solutions in subsequent steps (red box).

This approach is similar to alternative projection-based methods, i.e., mode superposition, Ritz vectors method \cite{Yoon20101744}, or the proper orthogonal decomposition \cite{Chatterjee2000}. The original problem is solved for  different values of the considered parameters in an offline stage and the pre-calculated solutions  are then collected in order to  construct a subspace.  However,  it is difficult to construct  such parameter spaces in a topological shape optimization problem. The number of parameters needs to be large enough to represent all possibilities of shape or topology changes, which makes the reduced solutions inaccurate or difficult to obtain. Instead of using the offline strategy, the stored displacements  from  consecutive iterations during the optimization process are used here to construct the subspace, and a reduced solution obtained without solving the full system in the subsequent iteration is expected to satisfy the equilibrium equation.

\section{Sensitivity analysis}\label{sect_sensitivity_analysis_reduced_model}
Due to the gradient-based optimization algorithm which is used, it is necessary to find the derivatives of the objective function and constraints with respect to the design variables $\alpha_{I}$. We follow the shape design sensitivity analysis by Choi et al. \cite{Choi2006structural}. The shape boundary of the physical domain is considered as the design variable, and shape sensitivity analyzes the relation of shape boundary perturbations to the objective function or constraints. By representing the shape boundary by a zero level set function, the perturbations of the shape boundary become a dynamic evolution of the level set function in time. The derivative of a domain functional with respect to shape can be replaced by the time derivative \cite{Wang2003227,Allaire2004363}.  In order to include the reduced solutions to the optimization algorithm, we define a general objective function  for each component of the homogenized elastic tensor as  follows
\begin{equation}\label{equ_object_func_with_reduced_model}
{\renewcommand{\arraystretch}{1.0}
    \begin{array}{*3{>{\displaystyle}l}} 
    J(\tilde{\pmb{\chi}},\pmb{\chi},\mathbf{v},\pmb{\lambda},\phi)= f(\tilde{\pmb{\chi}},\phi) + \left(a(\tilde{\pmb{\chi}},\mathbf{v},\phi) - l(\mathbf{v},\phi)\right) + \sum\limits_{m}^{nb}\left( a(\pmb{\chi}_{m},\pmb{\lambda}_{m},\phi) -l(\pmb{\lambda}_{m},\phi)\right),
    \end{array}
}
\end{equation}
where 
\begin{equation}\label{}
{\renewcommand{\arraystretch}{1.0}
    \begin{array}{*3{>{\displaystyle}l}} 
    f(\tilde{\pmb{\chi}},\phi)&=\int\limits_{Y} (\varepsilon^{0,ij}_{pq} - \varepsilon_{pq}(\tilde{\pmb{\chi}}^{ij})) C_{pqrs} (\varepsilon^{0,kl}_{rs} - \varepsilon_{rs}(\tilde{\pmb{\chi}}^{kl}))\mathcal{H}(\phi)  \, d\Omega ,    
    \end{array} 
}
\end{equation}
and variational problems in the reduced sub-space $\tilde{V}^{h}_{0}$ and in the finite dimensional space $V^{h}_{0}$ are defined respectively as
\begin{equation}\label{}
{\renewcommand{\arraystretch}{1.0}
    \begin{array}{*3{>{\displaystyle}l}} 
    a(\tilde{\pmb{\chi}},\mathbf{v},\phi)&= \int\limits_{Y} \varepsilon_{pq}(\tilde{\pmb{\chi}}^{ij}) C_{pqrs}\mathcal{H}(\phi) \, \varepsilon_{rs}(\mathbf{v})  \, d\Omega \\\\
    l(\mathbf{v},\phi)&= \int\limits_{Y} \varepsilon^{0,ij}_{pq}  C_{pqrs}\mathcal{H}(\phi) \, \varepsilon_{rs}(\mathbf{v})  \, d\Omega \\\\
    a(\pmb{\chi}_{m},\pmb{\lambda}_{m},\phi)&= \int\limits_{Y} \varepsilon_{pq}(\pmb{\chi}^{ij}_{m}) C_{pqrs}\mathcal{H}(\phi) \, \varepsilon_{rs}(\pmb{\lambda}_{m})  \, d\Omega \\\\
    l(\pmb{\lambda}_{m},\phi)&= \int\limits_{Y} \varepsilon^{0,ij}_{pq}  C_{pqrs}\mathcal{H}(\phi) \, \varepsilon_{rs}(\pmb{\lambda}_{m})  \, d\Omega .
    \end{array} 
}
\end{equation}
The subscript $m$ indicates the full solutions $\pmb{\chi}_{m}$ were used to construct reduced basis $\pmb{\phi}_{m}$. Shape derivatives of the objective function are given as
\begin{equation}\label{equ_derivative_object_with_RM}
{\renewcommand{\arraystretch}{1.0}
    \begin{array}{*3{>{\displaystyle}l}} 
    \dot{J} =  &f(\dot{\tilde{\pmb{\chi}}},\phi) + f(\tilde{\pmb{\chi}},\dot{\phi}) + a(\dot{\tilde{\pmb{\chi}}},\mathbf{v},\phi) + a(\tilde{\pmb{\chi}},\dot{\mathbf{v}},\phi) + a(\tilde{\pmb{\chi}},\mathbf{v},\dot{\phi}) - l(\dot{\mathbf{v}},\phi) - l(\mathbf{v},\dot{\phi}) \\\\
    & + \sum\limits_{m}^{nb}\left( a(\dot{\pmb{\chi}}_{m},\pmb{\lambda}_{m},\phi)+ a(\pmb{\chi}_{m},\dot{\pmb{\lambda}}_{m},\phi) +  a(\pmb{\chi}_{m},\pmb{\lambda}_{m},\dot{\phi}) -l(\dot{\pmb{\lambda}}_{m},\phi) - l(\pmb{\lambda}_{m},\dot{\phi})\right),
    \end{array}
}
\end{equation}
where the terms containing time derivatives, for example of the homogenized elastic tensor, have forms as
\begin{equation}\label{}
{\renewcommand{\arraystretch}{1.0}
    \begin{array}{*3{>{\displaystyle}l}} 
    f(\tilde{\pmb{\chi}},\dot{\phi}) = \int\limits_{Y} (\varepsilon^{0,ij}_{pq} - \varepsilon_{pq}(\tilde{\pmb{\chi}}^{ij})) C_{pqrs} (\varepsilon^{0,kl}_{rs} - \varepsilon_{rs}(\tilde{\pmb{\chi}}^{kl})) \, \delta(\phi) \, |\nabla \phi| V_{n}  \, d\Omega, \\\\
    f(\dot{\tilde{\pmb{\chi}}},\phi) = \int\limits_{Y} -2\,\varepsilon_{pq}(\dot{\tilde{\pmb{\chi}}}^{ij}) C_{pqrs} (\varepsilon^{0,kl}_{rs} - \varepsilon_{rs}(\tilde{\pmb{\chi}}^{kl})) \, \mathcal{H}(\phi)  \, d\Omega,
    \end{array}
}
\end{equation}
$\delta(\circ)=\mathcal{H}'(\circ)$ being the Dirac delta function. To avoid calculating the time derivative of the characteristic displacement $\dot{\tilde{\pmb{\chi}}}$ in \eqref{equ_derivative_object_with_RM}, the following adjoint equation is established and solved for $\mathbf{v}$ in general.
\begin{equation}\label{equ_adjoin_equation}
{\renewcommand{\arraystretch}{1.0}
    \begin{array}{*3{>{\displaystyle}l}} 
    f(\dot{\tilde{\pmb{\chi}}},\phi) + a(\dot{\tilde{\pmb{\chi}}},\mathbf{v},\phi) =0.
    \end{array}
}
\end{equation}
In fact \eqref{equ_adjoin_equation} is a self-adjoint problem and is satisfied if the test function $\mathbf{v}$ is properly selected such that
\begin{equation}\label{equ_self_adjoin_solution}
{\renewcommand{\arraystretch}{1.0}
    \begin{array}{*3{>{\displaystyle}l}} 
    \varepsilon_{rs}(\mathbf{v}) =2 \left(\varepsilon^{0,kl}_{rs} - \varepsilon_{rs}(\tilde{\pmb{\chi}}^{kl})\right).
    \end{array}
}
\end{equation}
All the terms containing $\dot{\pmb{\lambda}}_{m}$ can be eliminated since the following  relations hold true
\begin{equation}\label{}
{\renewcommand{\arraystretch}{1.0}
    \begin{array}{*3{>{\displaystyle}l}} 
    a(\pmb{\chi}_{m},\dot{\pmb{\lambda}}_{m},\phi) =l(\dot{\pmb{\lambda}}_{m},\phi) \quad \forall \pmb{\lambda}_{m} \in V^{h}_{0}, \quad m=1,...,nb.
    \end{array}
}
\end{equation}
If the standard solution is used instead of reduced solutions, i.e., $\tilde{\pmb{\chi}} \rightarrow \pmb{\chi}$, the residual is defined by 
\begin{equation}\label{equ_residual_equal_to_zero}
{\renewcommand{\arraystretch}{1.0}
    \begin{array}{*3{>{\displaystyle}l}} 
    r(\tilde{\pmb{\chi}},\dot{\mathbf{v}},\phi)=a(\tilde{\pmb{\chi}},\dot{\mathbf{v}},\phi) - l(\dot{\mathbf{v}},\phi) = 0, \quad \forall \dot{\mathbf{v}} \in V^{h}_{0}.
    \end{array}
}
\end{equation}
In this case, shape derivatives of the homogenized elastic tensor are 
\begin{equation}\label{}
{\renewcommand{\arraystretch}{1.0}
    \begin{array}{*3{>{\displaystyle}l}} 
    \frac{\partial J}{\partial \tau} = \frac{\partial C^{H}_{ijkl}}{\partial \tau} = - \int\limits_{Y} (\varepsilon^{0,ij}_{pq} - \varepsilon_{pq}(\pmb{\chi}^{ij})) C_{pqrs} (\varepsilon^{0,kl}_{rs} - \varepsilon_{rs}(\pmb{\chi}^{kl})) \delta(\phi) \, |\nabla \phi| V_{n}  \, d\Omega .
    \end{array}
}
\end{equation}
If a reduced solution $\tilde{\pmb{\chi}}$ is used, the residual is non-zero and written as
\begin{equation}\label{}
{\renewcommand{\arraystretch}{1.0}
    \begin{array}{*3{>{\displaystyle}l}} 
    r(\tilde{\pmb{\chi}},\dot{\mathbf{v}},\phi) = r(\tilde{\pmb{\chi}},\sum\limits_{m}^{nb} \varphi_{m}\dot{\pmb{\theta}}_{m},\phi)=\sum\limits_{m}^{nb} \varphi_{m} \, r(\tilde{\pmb{\chi}},\dot{\pmb{\theta}}_{m},\phi).
    \end{array}
}
\end{equation}
To eliminate the quantity  $\dot{\tilde{\pmb{\chi}}}_{m}$ in \eqref{equ_derivative_object_with_RM}, the following adjoint equations are formed for each set of the balance equations 
\begin{equation}\label{}
{\renewcommand{\arraystretch}{1.0}
    \begin{array}{*3{>{\displaystyle}l}} 
    a(\dot{\pmb{\chi}}_{m},\pmb{\lambda}_{m},\phi) +   \varphi_{m} \, r(\tilde{\pmb{\chi}},\dot{\pmb{\theta}}_{m},\phi) =0, \quad m=1,...,nb.
    \end{array}
}
\end{equation}
As the global ansatz function $\theta_{m}$ is constructed from the standard solutions $\chi$ in the finite-dimensional space $V^{h}_{0}$ (see \autoref{equ_global_ansatz_function}, \eqref{equ_contructed_RM01}-\eqref{equ_contructed_RM02}) and as the bilinear  form $a(\circ,\circ)$ is symmetric in its arguments, the adjoint equation finally has the form as
\begin{equation}\label{equ_adjoint_equation_lamda}
{\renewcommand{\arraystretch}{1.0}
    \begin{array}{*3{>{\displaystyle}l}} 
    a(\pmb{\lambda}_{m},\dot{\pmb{\chi}}_{m},\phi) +   \varphi_{m} \, r(\tilde{\pmb{\chi}},\dot{\pmb{\chi}}_{m},\phi) =0, \quad m=1,...,nb
    \end{array}
}
\end{equation}
and solved for the adjoint variables $\pmb{\lambda}_{m}$. The final expression of the sensitivity is
\begin{equation}\label{equ_final_time_derivative_effective_tensor_00}
{\renewcommand{\arraystretch}{1.0}
    \begin{array}{*3{>{\displaystyle}l}} 
    \dot{J} =  &  f(\tilde{\pmb{\chi}},\dot{\phi})   + a(\tilde{\pmb{\chi}},\mathbf{v},\dot{\phi})  - l(\mathbf{v},\dot{\phi}) 
    + \sum\limits_{m}^{nb}\left(     a(\pmb{\chi}_{m},\pmb{\lambda}_{m},\dot{\phi})  - l(\pmb{\lambda}_{m},\dot{\phi})\right)
    \end{array}
}
\end{equation}
in which adjoint variables $\mathbf{v}$ and $\pmb{\lambda}_{m}$ are obtained from the self-adjoint problem \eqref{equ_self_adjoin_solution} and \eqref{equ_adjoint_equation_lamda} respectively. If the Hamilton-Jacobi equation is replaced by an original differential equation \eqref{equ_H_J_ODE}, any integral function involving the normal velocity component 
\begin{equation}\label{}
{\renewcommand{\arraystretch}{1.0}
    \begin{array}{*3{>{\displaystyle}l}} 
    \dot{J}=\frac{\partial J(\circ, \phi)}{\partial \tau}=g(\circ,\dot{\phi}) = \int\limits_{Y}  (\circ) \, \delta(\phi) |\nabla \phi| V_{n}  \, d\Omega
    \end{array}
}
\end{equation}
can be written as
\begin{equation}\label{equ_final_time_derivative_effective_tensor_01}
{\renewcommand{\arraystretch}{1.0}
    \begin{array}{*3{>{\displaystyle}l}} 
    \frac{\partial J(\circ, \phi)}{\partial \tau}  =\sum\limits_{I=1}^{ncp} \left(\int\limits_{Y} (\circ) \delta(\phi) R_{I}(\mathbf{x}) \, d\Omega \right) \cdot  \dot{\alpha}_{I}.
    \end{array}
}
\end{equation}
Additionally, the time derivative of the objective function $J$ can be obtained by applying the chain rule
\begin{equation}\label{equ_chain_rule_effective_tensor} 
{\renewcommand{\arraystretch}{1.0}
    \begin{array}{*3{>{\displaystyle}l}} 
    \frac{\partial J(\circ, \phi)}{\partial \tau}  = \sum\limits_{I=1}^{ncp}   \frac{\partial J(\circ, \phi)}{\partial \alpha_{I}}  \, \frac{\partial \alpha_{I}}{\partial \tau} .
    \end{array}
}
\end{equation}
By comparing \eqref{equ_chain_rule_effective_tensor} and \eqref{equ_final_time_derivative_effective_tensor_01}, the sensitivity analysis of the objective function with respect to the design variable $\alpha_{I}$ is
\begin{equation}\label{}
{\renewcommand{\arraystretch}{1.0}
    \begin{array}{*3{>{\displaystyle}l}} 
    \frac{\partial J(\circ, \phi)}{\partial \alpha_{I}}  =  \int\limits_{Y} (\circ) \delta(\phi) R_{I}(\mathbf{x}) \, d\Omega .
    \end{array}
}
\end{equation}

The  sensitivity analysis  of volume constraint is  given as
\begin{equation}\label{}
{\renewcommand{\arraystretch}{1.0}
    \begin{array}{*3{>{\displaystyle}l}} 
    \frac{\partial V}{\partial \alpha_{I}} = \int\limits_{Y} \delta(\phi) R_{I}(\mathbf{x}) \, d\Omega.
    \end{array}
}
\end{equation}

\FloatBarrier
\section{Numerical examples}\label{sect_numerical_examples}
The homogenization method detailed in \cref{sect_homogenization_method} allows to predict the macroscopic properties of composites from a unit cell geometry. Designing the behavior of composite material is replaced by designing a unit cell such that the effective properties meet requirements. Seeking for auxetic materials turns into designing the effective elasticity tensor having negative Poisson's ratio. Therefore, we can assume that the target properties of material are given, and the optimization process is to search for the unit cell geometries such that the difference between effective properties and target properties is minimized. 
\begin{equation}\label{}
{\renewcommand{\arraystretch}{1.5}
    \begin{array}{*3{>{\displaystyle}l}} 
    \min & J_{1} =   \frac{1}{2} \sum\limits_{i,j,k,l=1}^{d} \omega_{ijkl} \times (C_{ijkl}-C^{\star}_{ijkl})^2 \\ 
    \text{s.t.}  &\left\{\begin{array}{*3{>{\displaystyle}l}} 
    a(\pmb{\chi},\mathbf{v},\phi) = l(\mathbf{v},\phi) \quad (\text{equilibrium equation})\\
    V(\phi) = \int\limits_{Y} \mathcal{H}(\phi)\, d\Omega \leq v_{f} \quad (\text{volume constraint})\\
    \alpha_{\min}\leq\alpha_{i}\leq\alpha_{\max} \quad (i=1,2,...,N) \quad (\text{design variable constraint})
    \end{array}\right.
    \end{array}
}
\end{equation}
where $C^{\star}_{ijkl}$ are the components of the target elasticity matrix which has the following form in the two-dimensional dimension
\begin{equation}\label{}
{\renewcommand{\arraystretch}{1.0}
    \begin{array}{*3{>{\displaystyle}l}} 
    \mathbf{C}^{\star} = \left[\begin{array}{ccc} 
    C^{\star}_{1111} & C^{\star}_{1122}& 0\\
    C^{\star}_{1122}& C^{\star}_{2222} & 0\\
    0 & 0 & C^{\star}_{1212}  \\  
    \end{array}\right],
    \end{array}
}
\end{equation}
and $\omega_{ijkl}$ are weighting factors used in the least-square function. The bilinear energy form $a(\pmb{\chi},\mathbf{v},\phi)$ and linear load form  $l(\mathbf{v},\phi)$ are defined in \eqref{equ_equilibrium_equation_in_unit_cell_with_lsf}.

We first consider 2D unit cells  subjected to plane stress conditions and   discretized with an IGA mesh with  $60\times 60$ order $p=2$ polynomial B{\'e}zier elements. The solid part in the unit cell is constituted by an isotropic material having  Young's modulus $E=1.0$ GPa and Poisson's ratio $\nu=0.3$. The target elasticity components are set to $C^{\star}_{1111}=C^{\star}_{2222}=0.1$, $C^{\star}_{1122}=-0.05$ which ensures that the final configuration has negative Poisson's ratio. We set the weighting factors $\omega_{1111}=\omega_{2222}=0.01$, $\omega_{1122}=0.5$. For a level set function initialized with expansion coefficients, the upper bound  $\alpha_{\max}=\max \{\alpha^{0}_{I}\}$ and lower bound  $\alpha_{\min}=\min\{\alpha^{0}_{I}\}$ are chosen in the subsequent iterations. We use the method of moving asymptotic (MMA) \cite{Svanberg1987} to update the design variables.  Orthotropic materials are obtained by considering the geometry symmetry of square unit cells, the expansion coefficients $\alpha_{I}$ in one-quarter of the unit cell are updated and the remaining configurations are obtained from the reflectional symmetry.

Optimal solutions corresponding to different initial configurations, target volumes are given in \autoref{fig_optimized_designs_from_different_initial_configs}. Despite the initial configurations consist of  simple circles, the complex shapes with sharp boundaries and thin ribs occur in the designs. There is no gray area or blur boundaries, the solid and void interfaces are clear. The verification or fabrication procedure can be directly performed for those final designs. These advantages of features of level set methods are highly preferable.  Final shapes of unit cells are different due to the fact that the optimization algorithm (MMA) is a local optimizer. To summarize, the generated unit cells with negative Poisson's ratios are obtained and similar to well-known auxetic structures: anti-tetra-chiral structures \cite{Li2017Antitetrachiral} (see \autoref{fig_optimized_designs_from_different_initial_configs}(c)),  re-entrant structures \cite{Masters1996403} (see \autoref{fig_optimized_designs_from_different_initial_configs}(b, d)), or auxetic microstructures obtained with SIMP \cite{Xia2015} (see \autoref{fig_optimized_designs_from_different_initial_configs}(a, e)). 

The volume fraction constraints significantly affect the final designs. In fact, negative Poisson's ratio in materials is created by a mechanism with trusses or beams connected through hinges. These structures induce perpendicular motions among components by bending or rotating at the hinges. As much material is removed, a higher possibility of thin layers is produced. The amount of material removed also drives elastic stiffness of unit cells, this is a trade-off between the effective Young's modulus and negative Poisson's ratios in auxetic microstructure optimization. Softer materials have higher auxetic effects and vice-visa.

Geometries of the unit cell are plotted as selected iterations shown in \autoref{fig_convergence_of_the_objective_function_without_rom} illustrating topology change during the optimization process. At the beginning, material is removed to satisfy the volume constraint and the objective function oscillates significantly within the first 20 iterations. Topology changes mainly take place in this material removal process. Dynamic moving of boundaries with merging and splitting allows to remove existing holes or create new holes, increasing the design flexibility of shape and topology. Solid domains link together and meaningless features like islands are eliminated while the objective function values are reducing gradually. After 125 iterations, the objective function is kept unchanged and Poisson's ratios are minimized.

We next investigate the influence of the reduced basis approach to  the convergence rate and  final design. The same initial configuration is considered but the algorithm with reduced order model (as shown in \autoref{fig_optimization_algorithm_with_reduced_basis}) is used in this example. A reduced basis with size $nb=12$ is constructed and enriched during the optimization process. Reduced solutions are accepted and used in further calculation steps if the residual error is smaller than a threshold $tol=0.01$ for all test cases ($mn=11,22,12$) simultaneously. As shown in \autoref{fig_convergence_of_the_objective_function_with_rom}, the objective function varies dramatically in the first few iterations  along with the large change in shape and topology due to material removal. These imply that the reduced solutions, generated by a set of  different configurations, will not satisfy the condition \eqref{equ_error_reduce_solution}, and full solutions are required to be computed. When the objective function values are gradually decreasing, a set of similar geometries and corresponding basis vectors enrich the reduced basis sufficiently, this subspace enables to produce reduced solutions which are accurate enough to proceed with the next calculation steps. The blue squares indicate that the optimization algorithm applies the reduced model order successfully and computational burden of solving the original system is eliminated by inverting the systems with small size, i.e., $nb=12$.

\begin{center}
    \begin{tabular}{>{\centering\arraybackslash} m{2cm} >{\centering\arraybackslash} m{2.5cm} >{\centering\arraybackslash} m{3.25cm}  >{\centering\arraybackslash} m{3cm} >{\centering\arraybackslash} m{1.0 cm} }\hline
        \small Initial design & \small Optimized unit cell  &  \small $2\times 2$ repeated unit cells &  \small Effective property \& volume fraction & \\
        \hline\\
        
        (a) \includegraphics[width=1.0\linewidth]{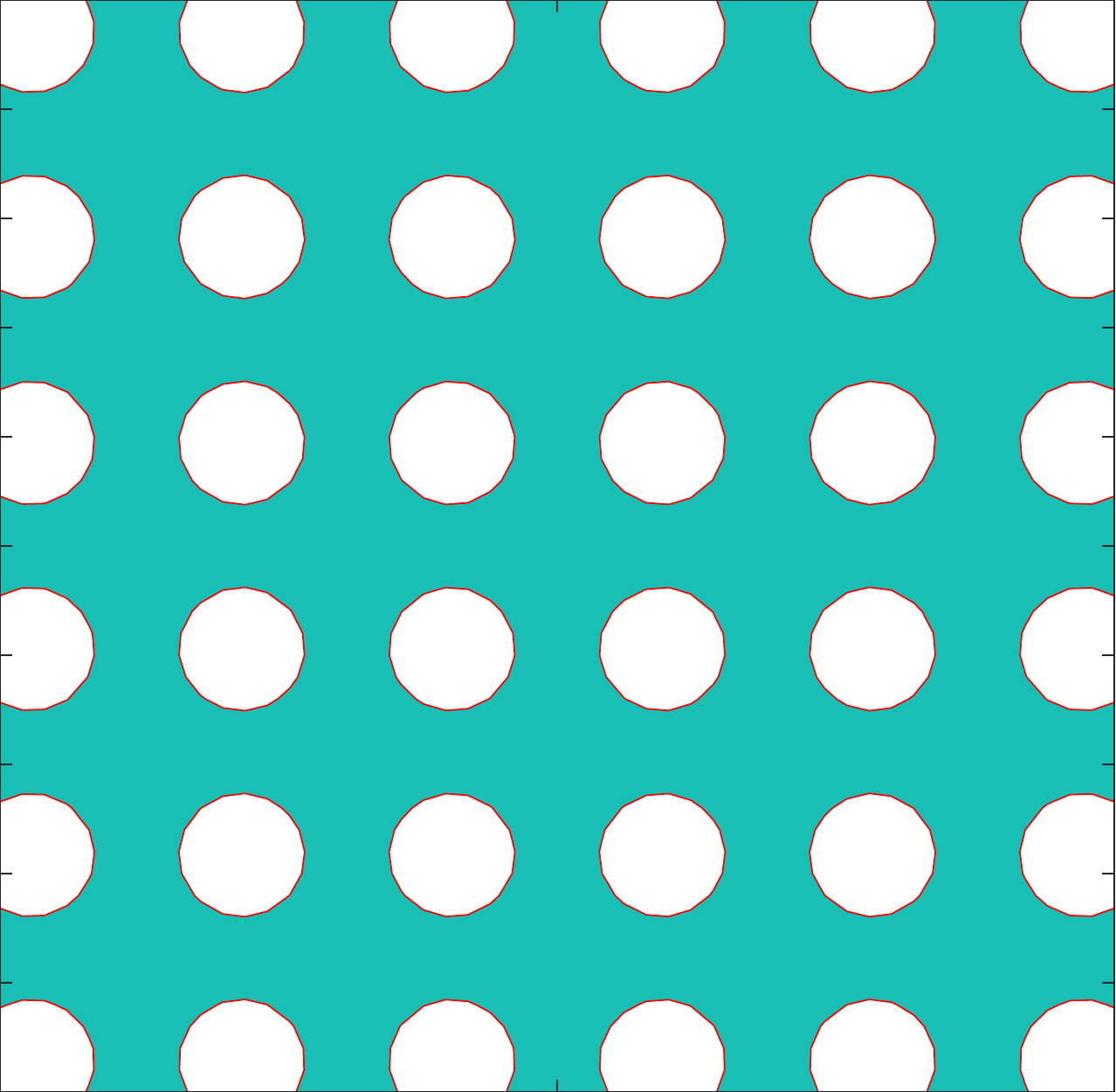} & \includegraphics[width=1.0\linewidth]{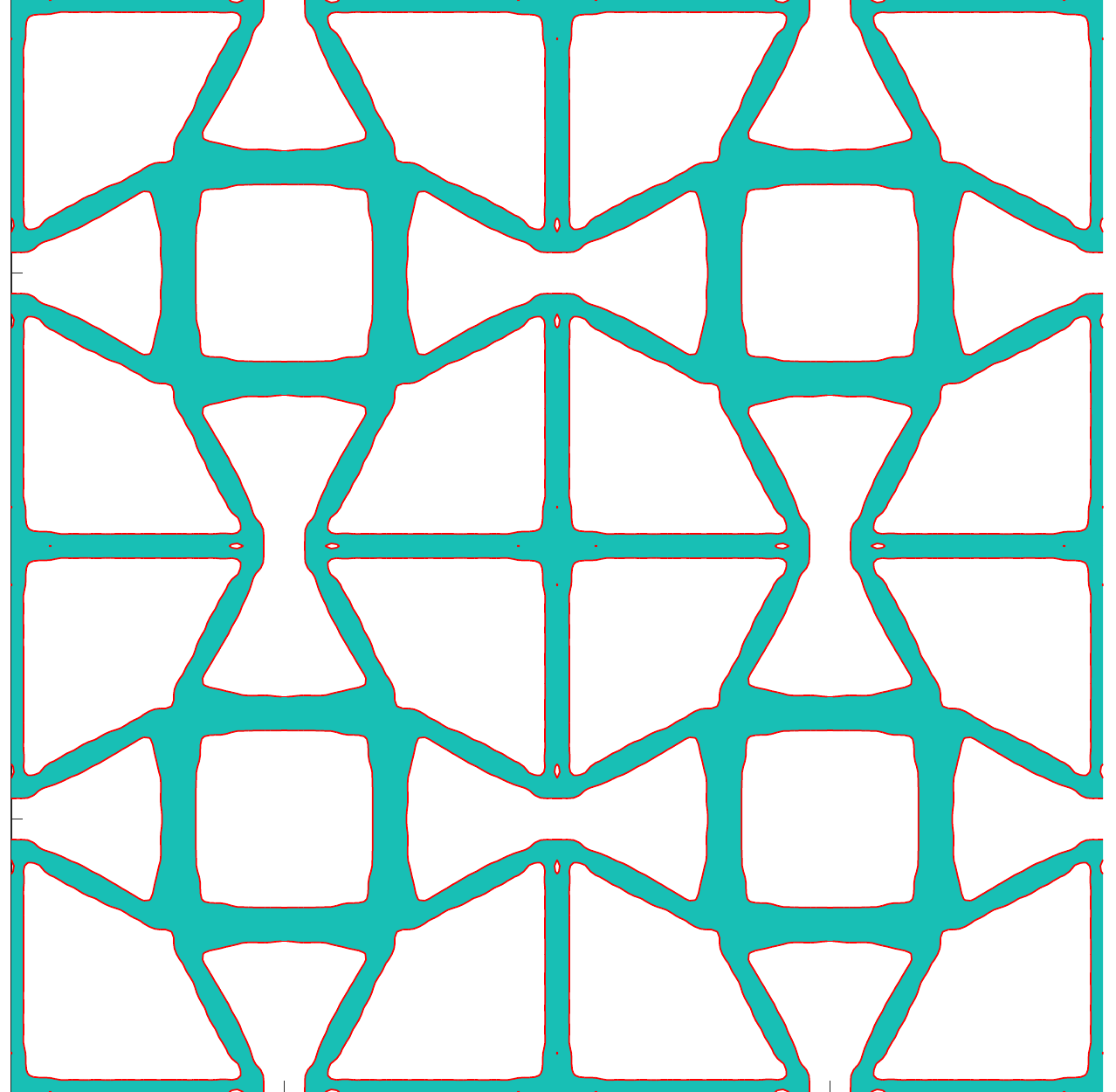} & \includegraphics[width=1.0\linewidth]{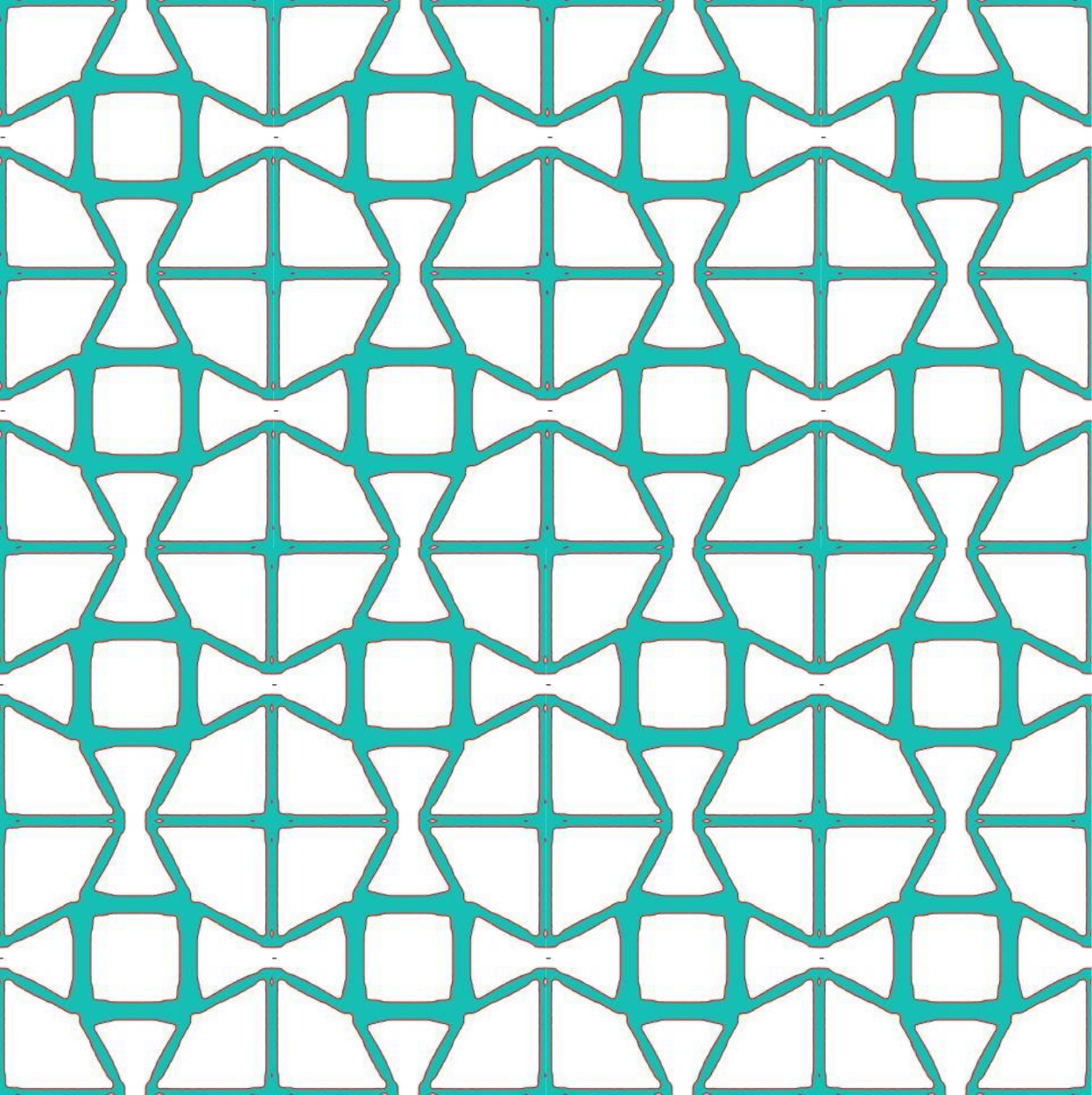}&  
        \scalebox{0.75}
        {
            $ 
            {\renewcommand{\arraystretch}{1.0}
                \begin{array}{*3{>{\displaystyle}l}} 
                \mathbf{C}^{H}=\left[\begin{array}{*3{>{\displaystyle}c}} 
                0.0658  & -0.0372 &0 \\
                -0.0372 & 0.0658 &0 \\
                0 & 0 &0.0020 
                \end{array}\right]\\\\
                V_{f}=0.275\\\\
                \nu_{xy}=\nu_{yx}=-0.565
                \end{array}
            }
            $
        } &\\
        \hline\\

        (b) \includegraphics[width=1.0\linewidth]{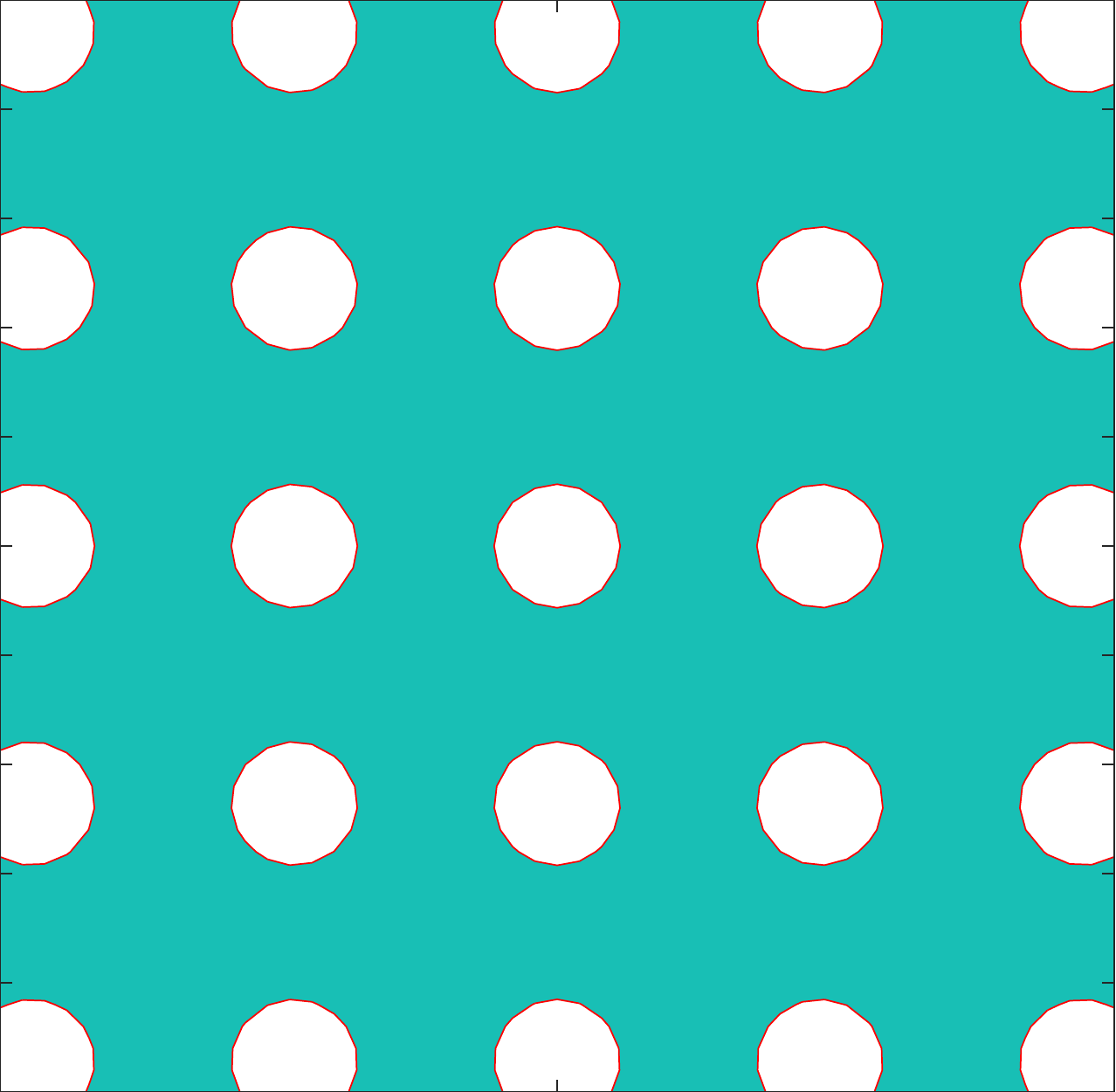} & \includegraphics[width=1.0\linewidth]{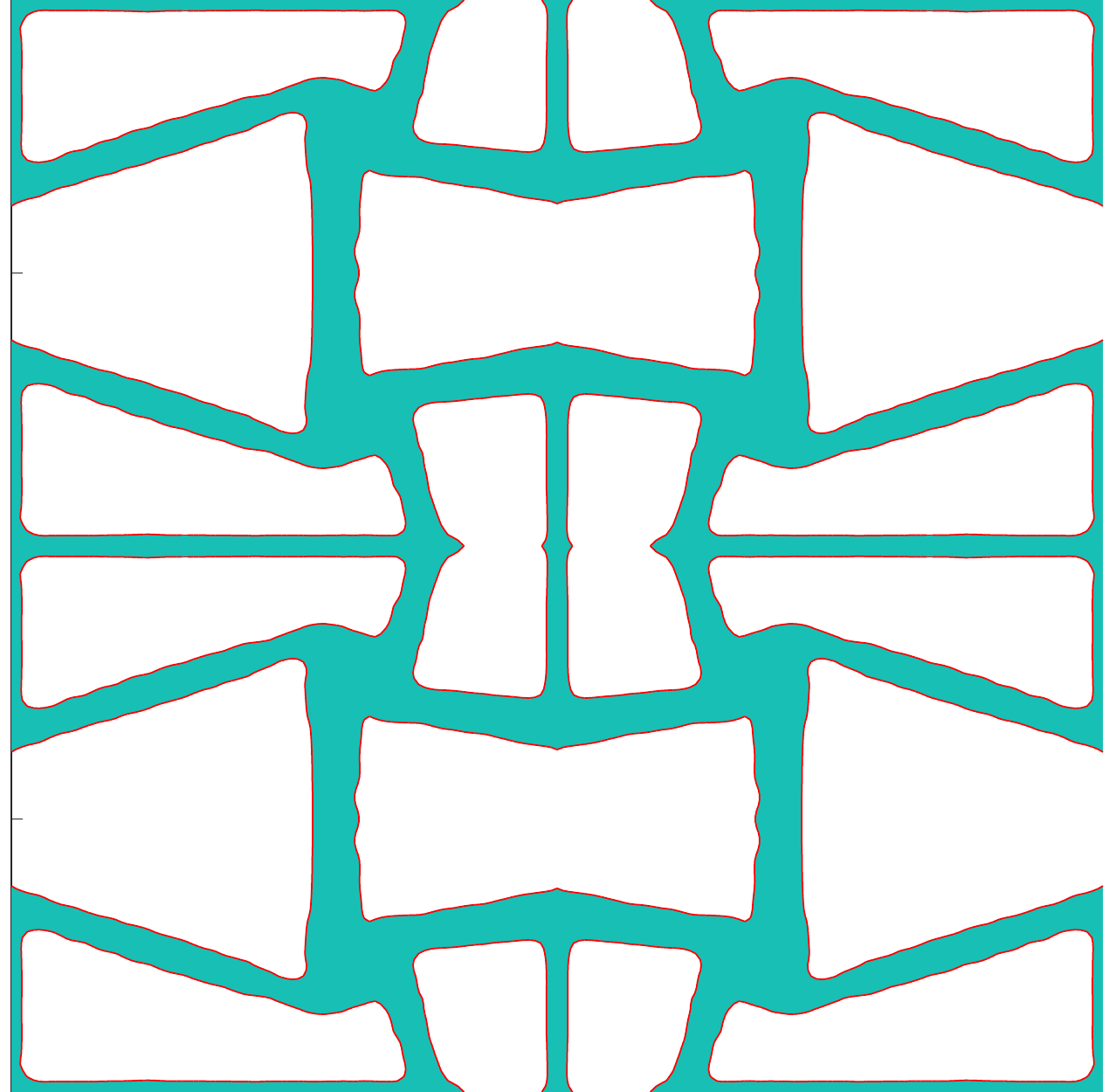} & \includegraphics[width=1.0\linewidth]{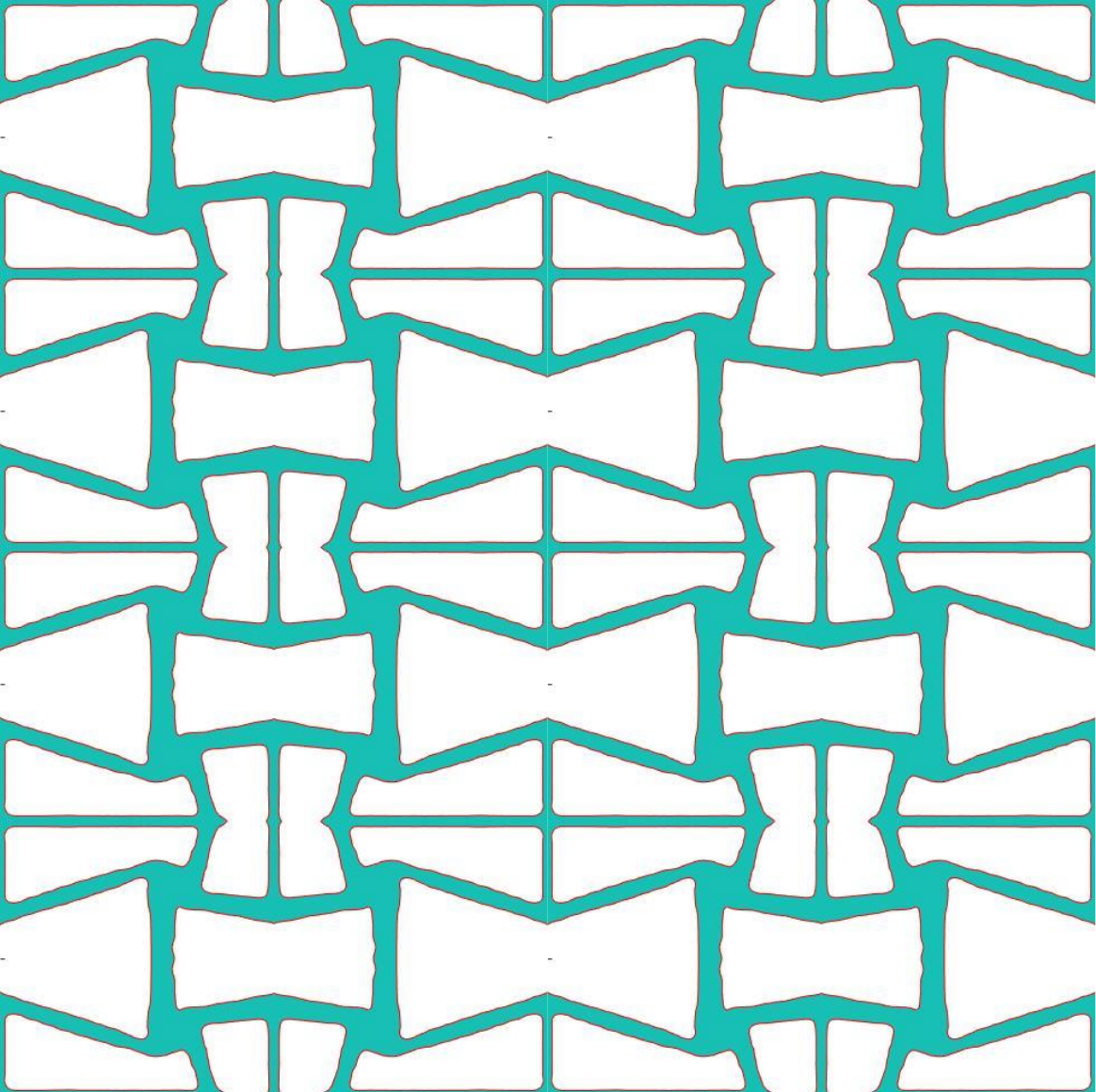}&  
        \scalebox{0.75}
        {
            $ 
            {\renewcommand{\arraystretch}{1.0}
                \begin{array}{*3{>{\displaystyle}l}} 
                \mathbf{C}^{H}=\left[\begin{array}{*3{>{\displaystyle}c}} 
                0.0921  & -0.0430 &0 \\
                -0.0430 & 0.05 &0 \\
                0 & 0 &0.0014
                \end{array}\right]\\\\
                V_{f}=0.275\\\\
                \nu_{xy}=-0.466, \, \nu_{yx}=-0.860
                \end{array}
            }
            $
        } &\\
        \hline\\
        
        (c) \includegraphics[width=1.0\linewidth]{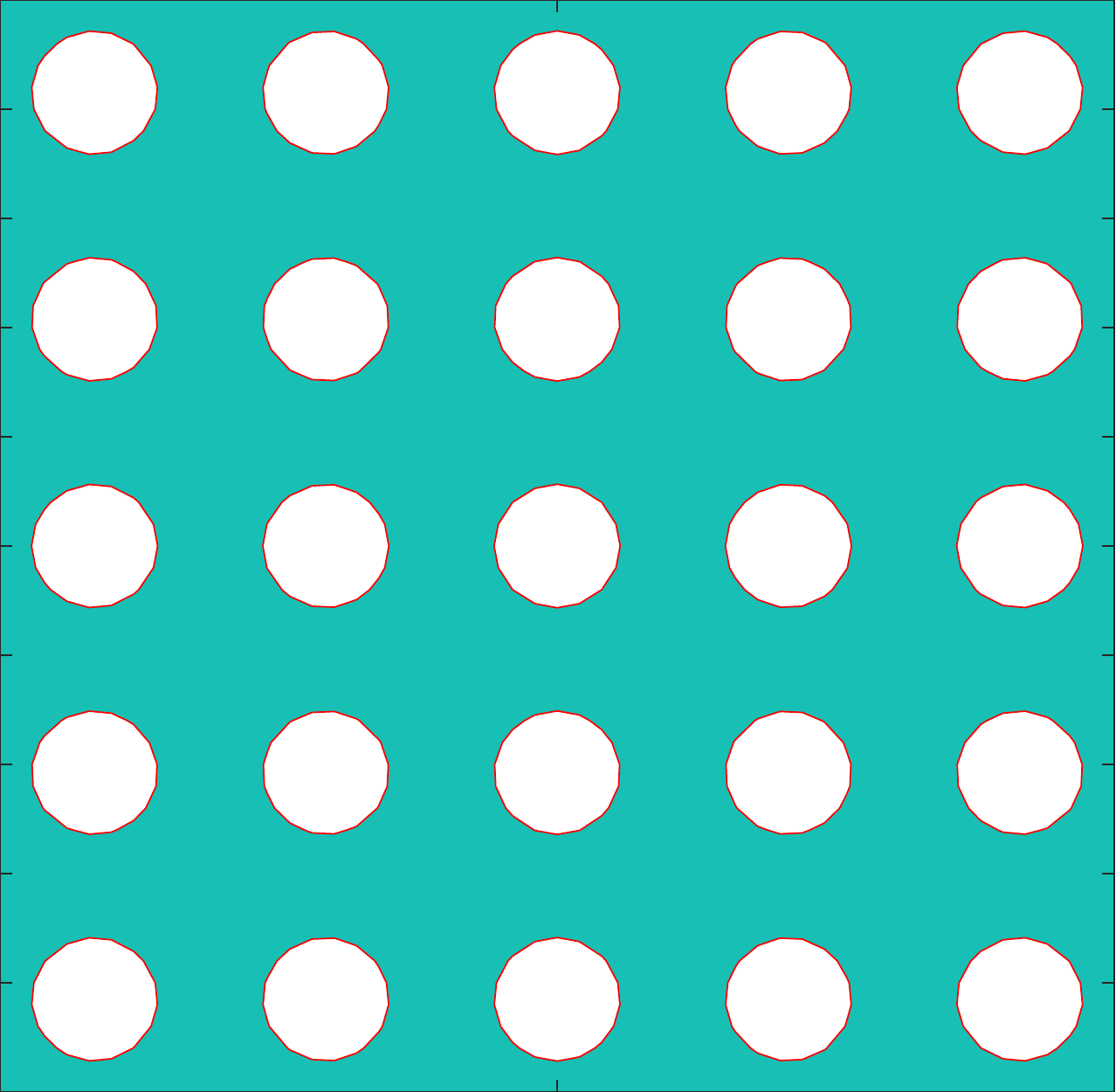} & \includegraphics[width=1.0\linewidth]{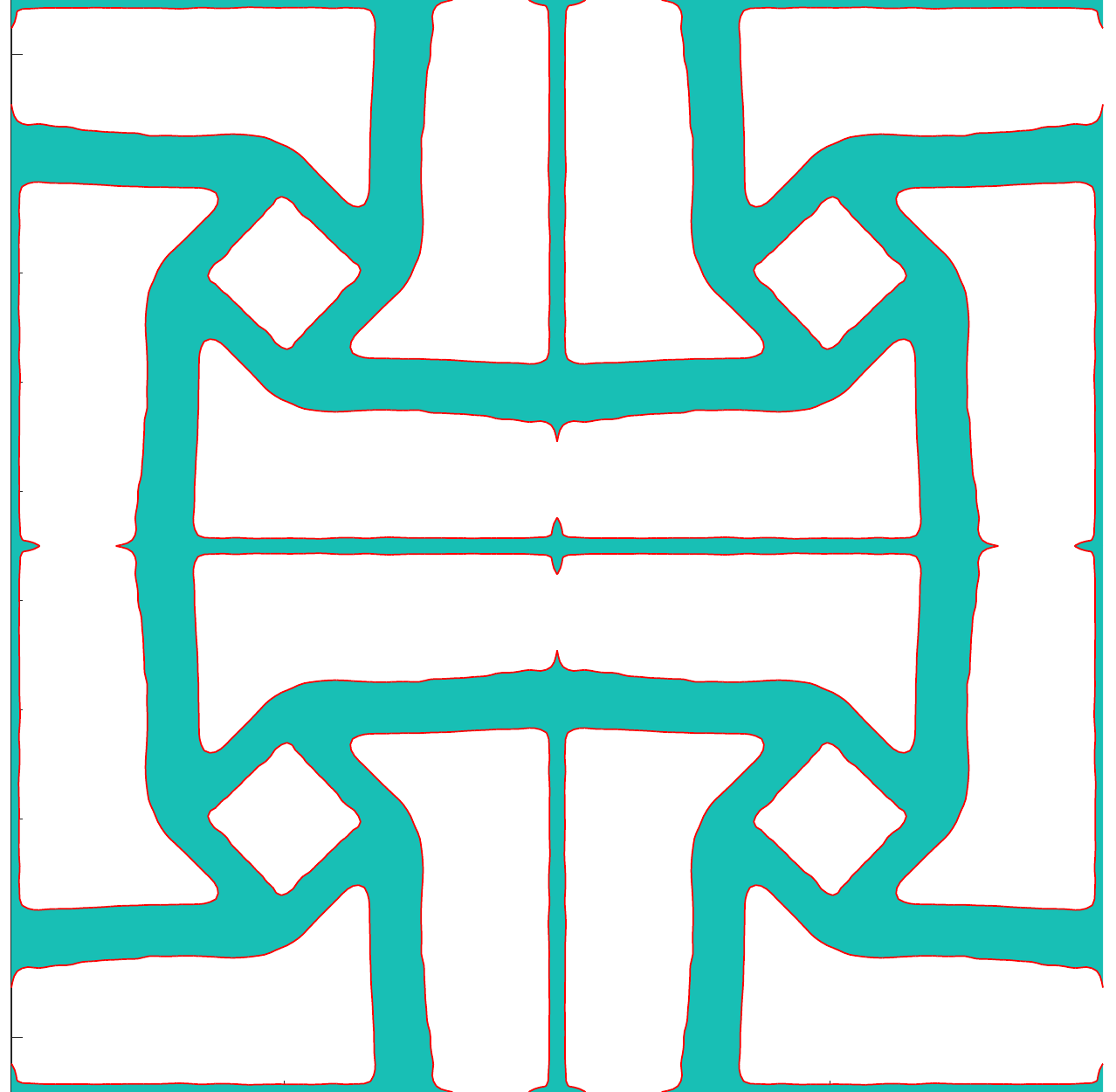} & \includegraphics[width=1.0\linewidth]{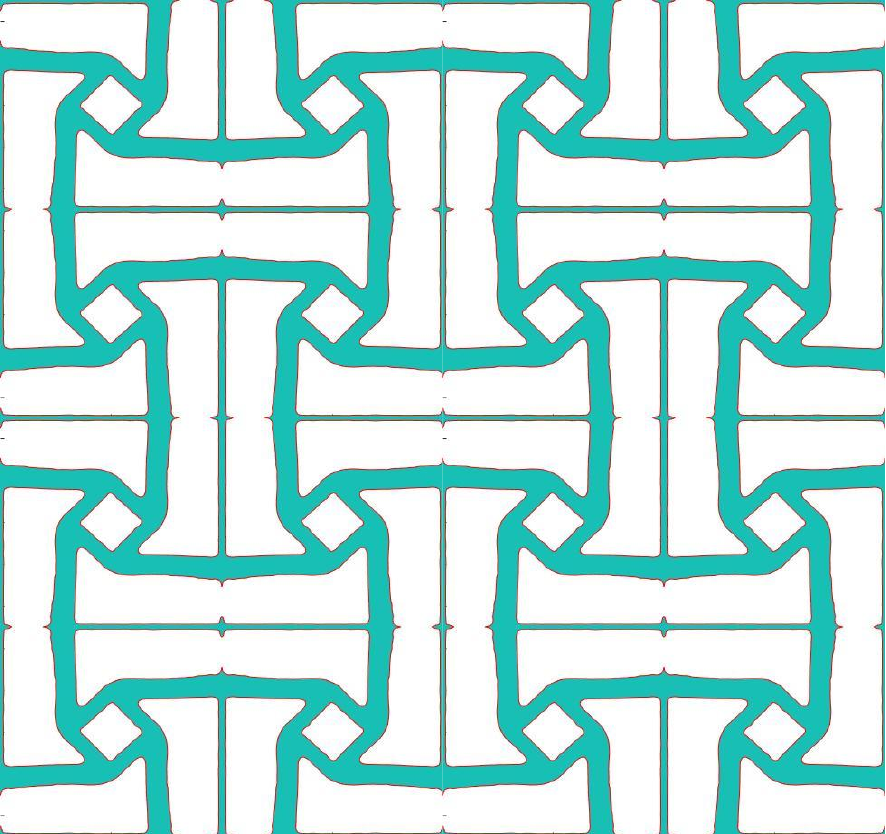}&  
        \scalebox{0.75}
        {
            $ 
            {\renewcommand{\arraystretch}{1.0}
                \begin{array}{*3{>{\displaystyle}l}} 
                \mathbf{C}^{H}=\left[\begin{array}{*3{>{\displaystyle}c}} 
                0.046  & -0.039 &0 \\
                -0.039 & 0.046 &0 \\
                0 & 0 &0.0012 
                \end{array}\right]\\\\
                V_{f}=0.3\\\\
                \nu_{xy}=\nu_{yx}=-0.830
                \end{array}
            }
            $
        } &\\
        \hline\\
        
        (d) \includegraphics[width=1.0\linewidth]{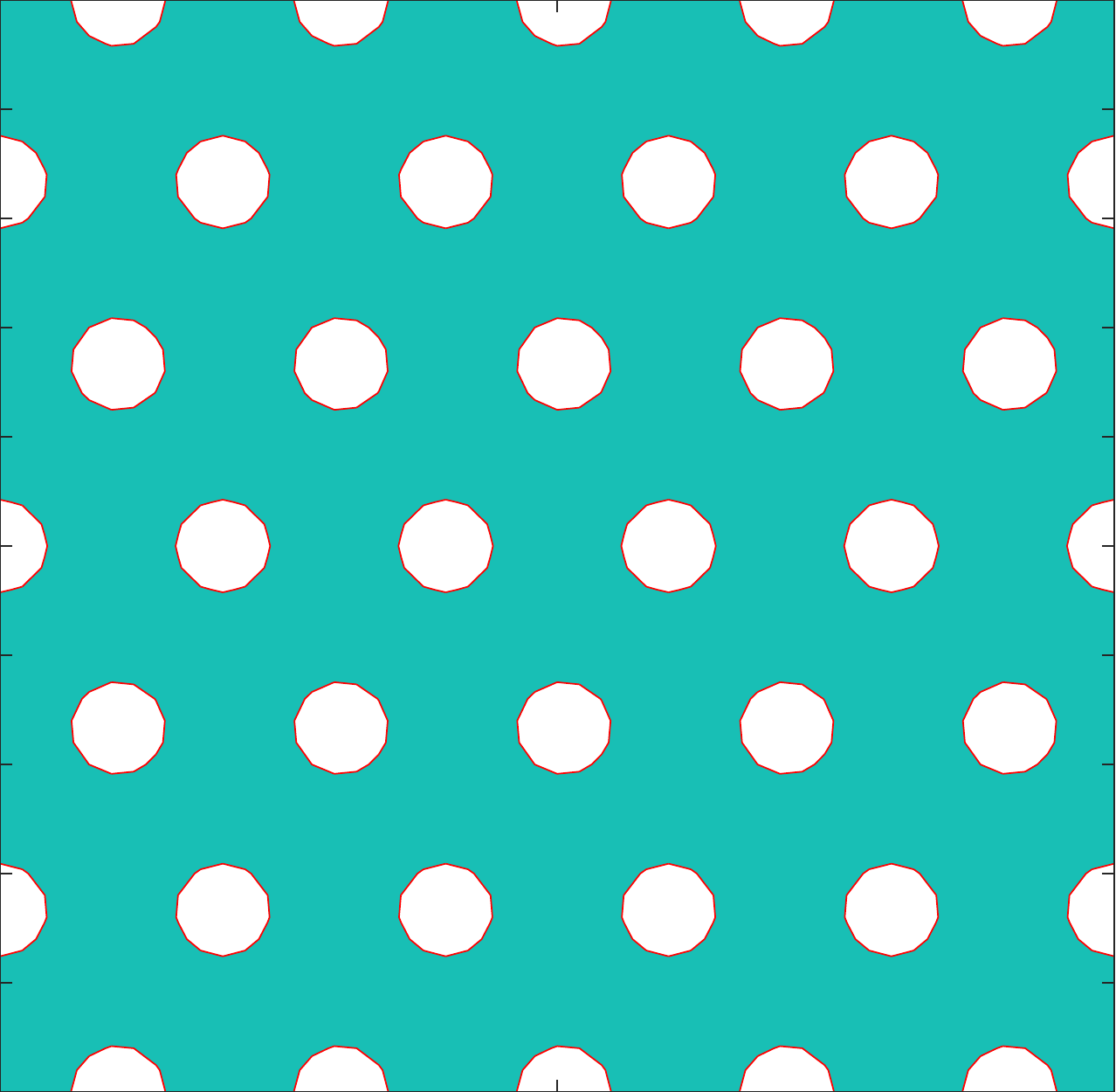} & \includegraphics[width=1.0\linewidth]{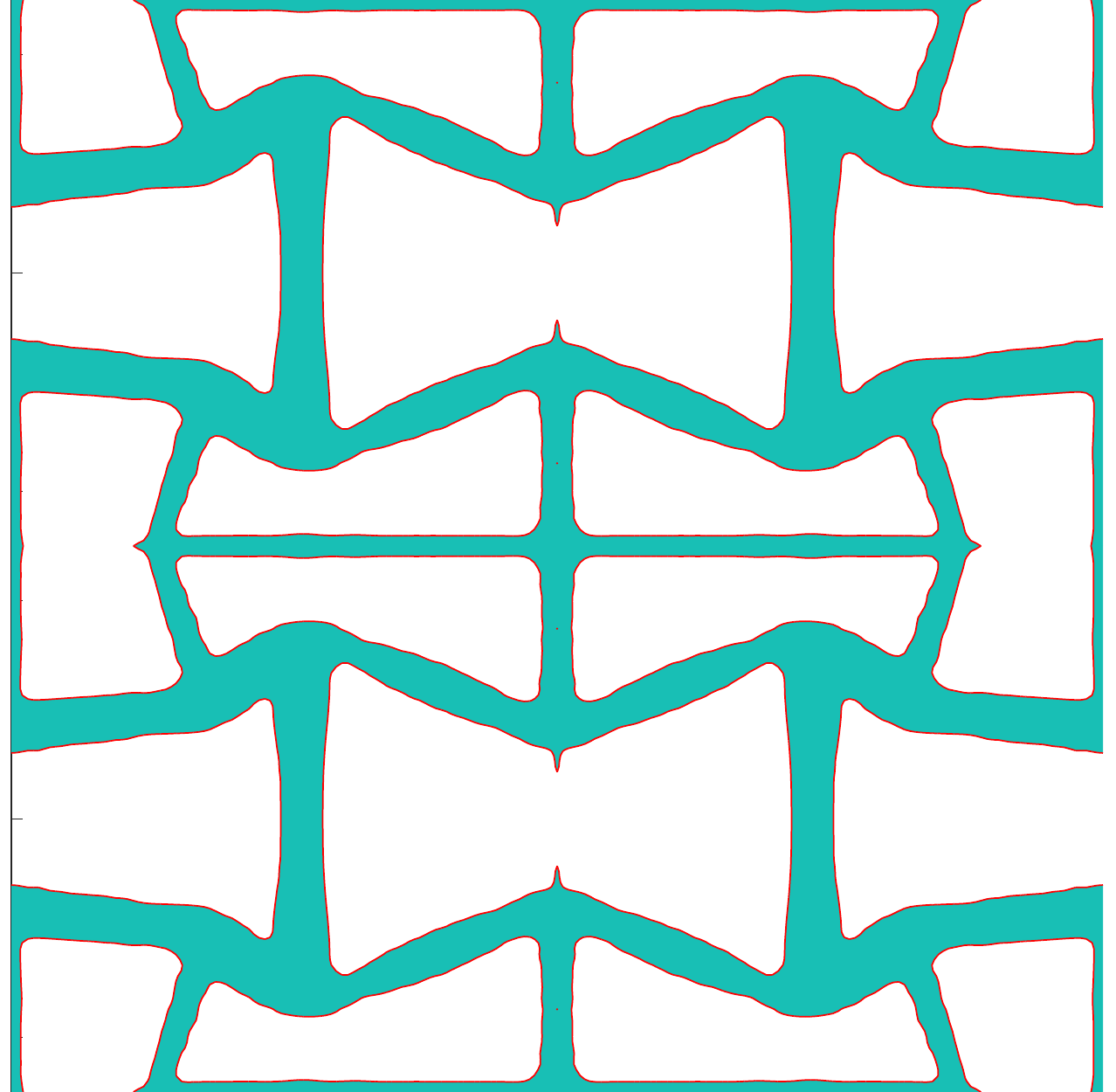} & \includegraphics[width=1.0\linewidth]{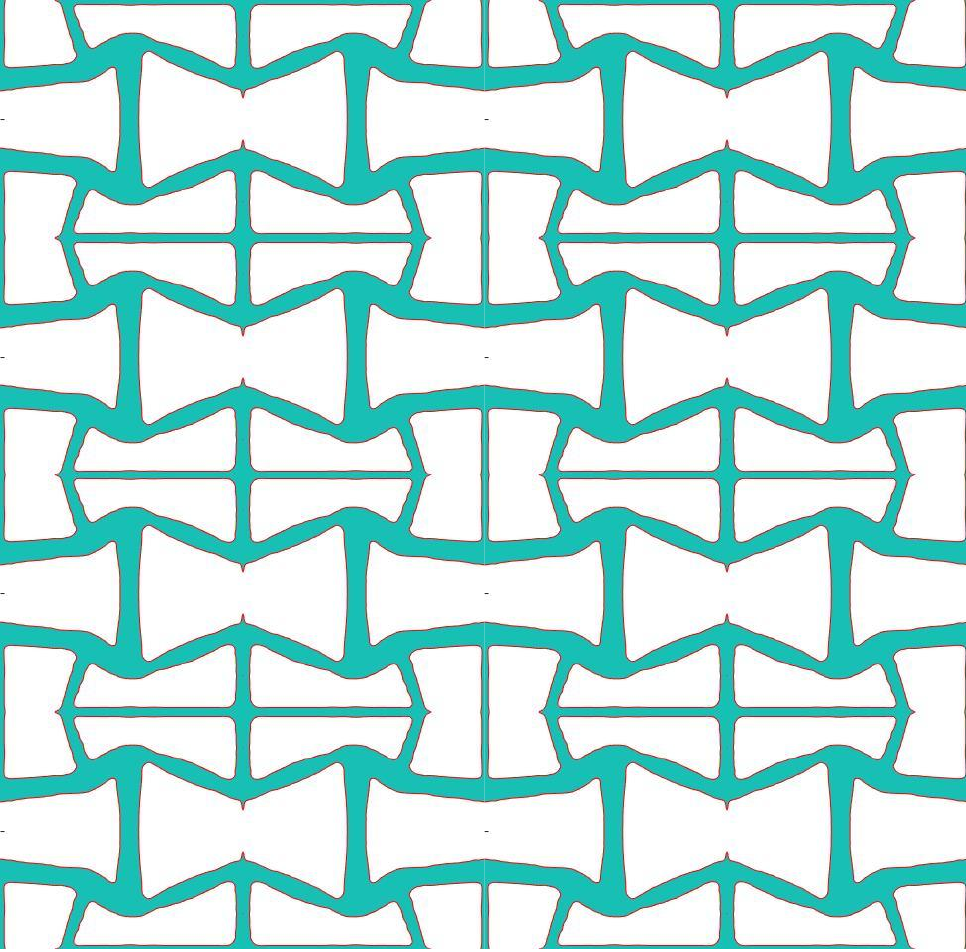}&  
        \scalebox{0.75}
        {
            $ 
            {\renewcommand{\arraystretch}{1.0}
                \begin{array}{*3{>{\displaystyle}l}} 
                \mathbf{C}^{H}=\left[\begin{array}{*3{>{\displaystyle}c}} 
                0.1094  & -0.0419 &0 \\
                -0.0419 & 0.052 &0 \\
                0 & 0 &0.0018 
                \end{array}\right]\\\\
                
                V_{f}=0.3 \\\\
                \nu_{xy}=-0.383, \, \nu_{yx}=-0.805
            \end{array}
        }
        $
    } &\\
    \hline\\

    (e) \includegraphics[width=1.0\linewidth]{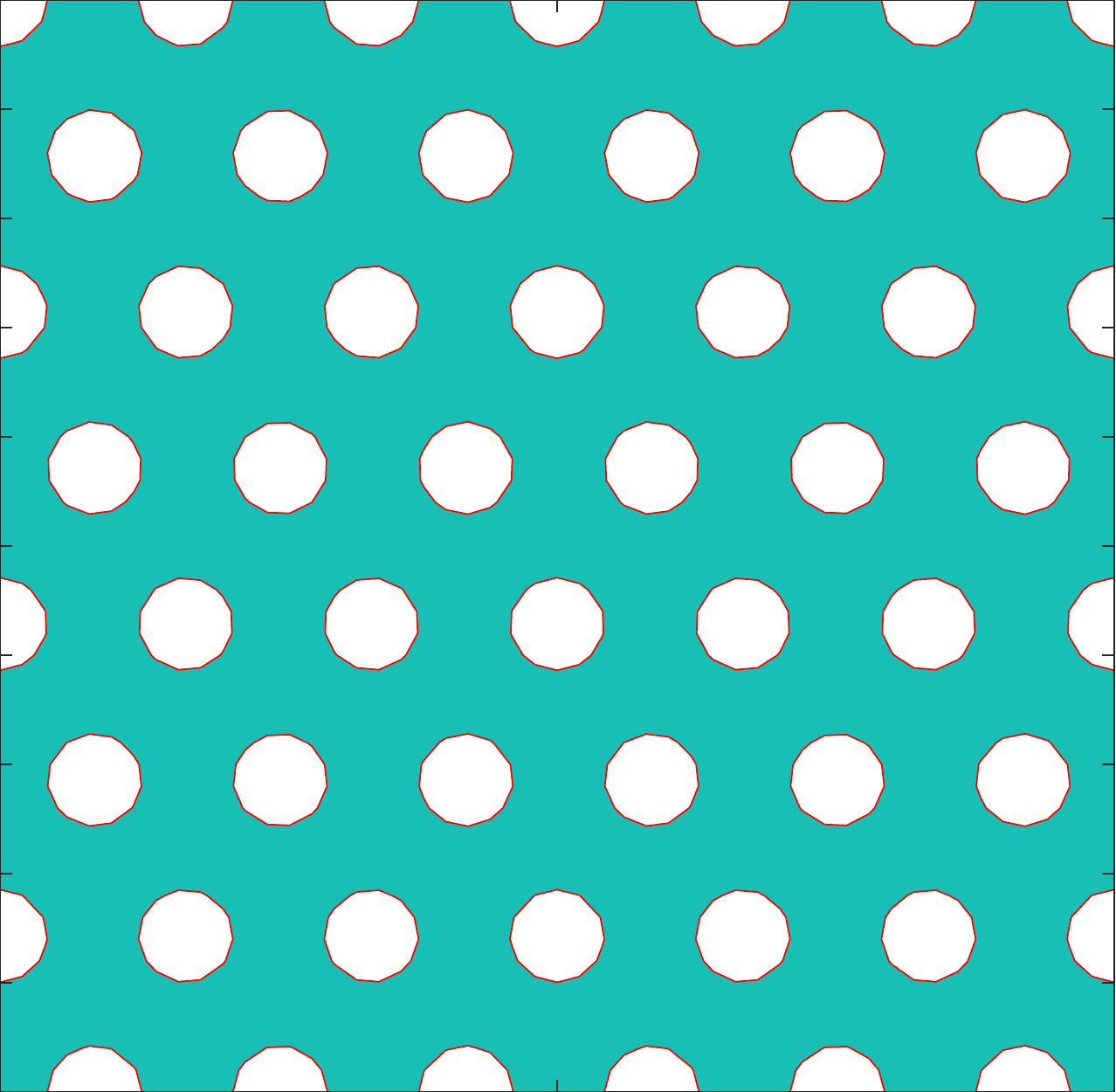} & \includegraphics[width=1.0\linewidth]{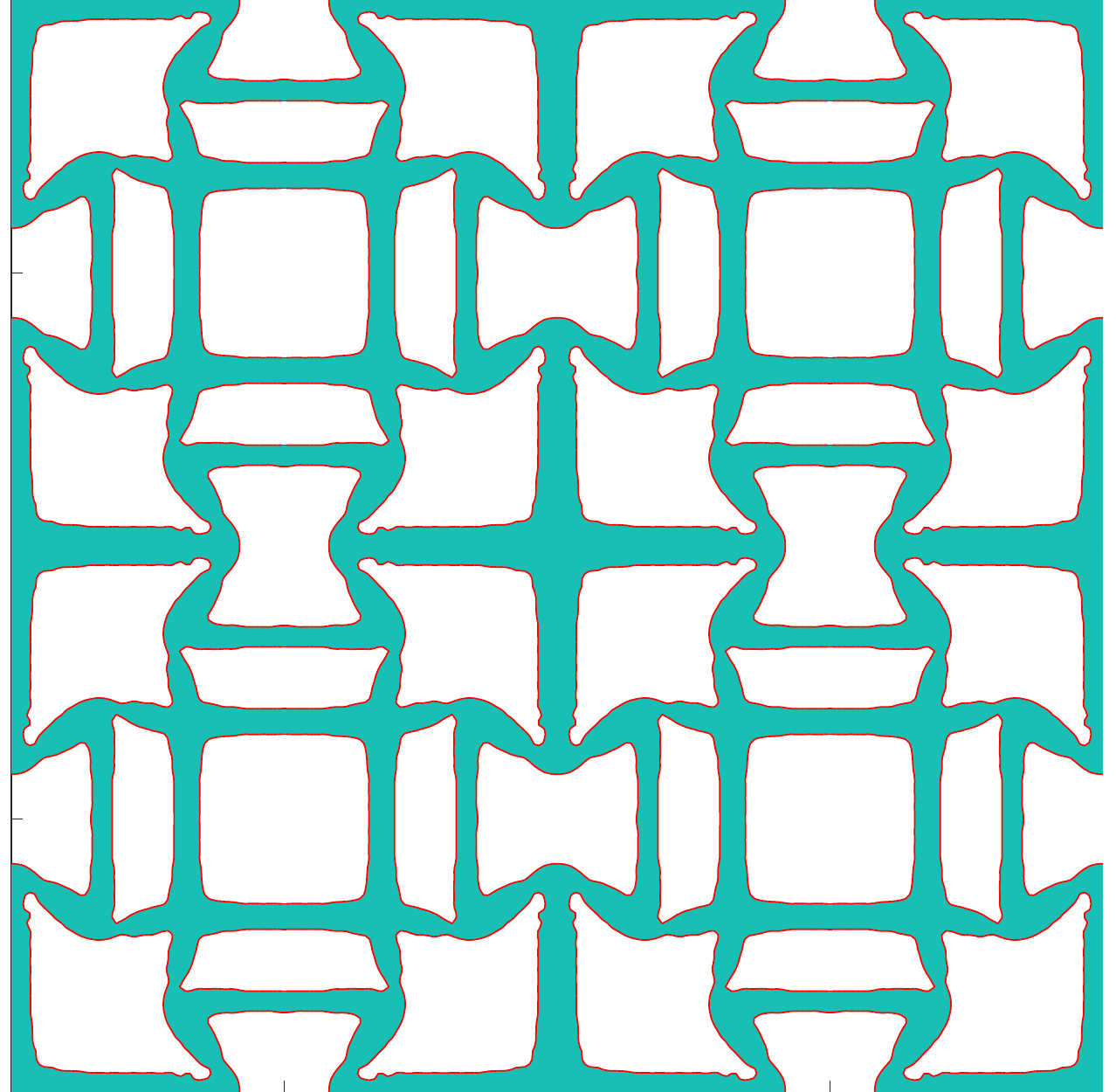} & \includegraphics[width=1.0\linewidth]{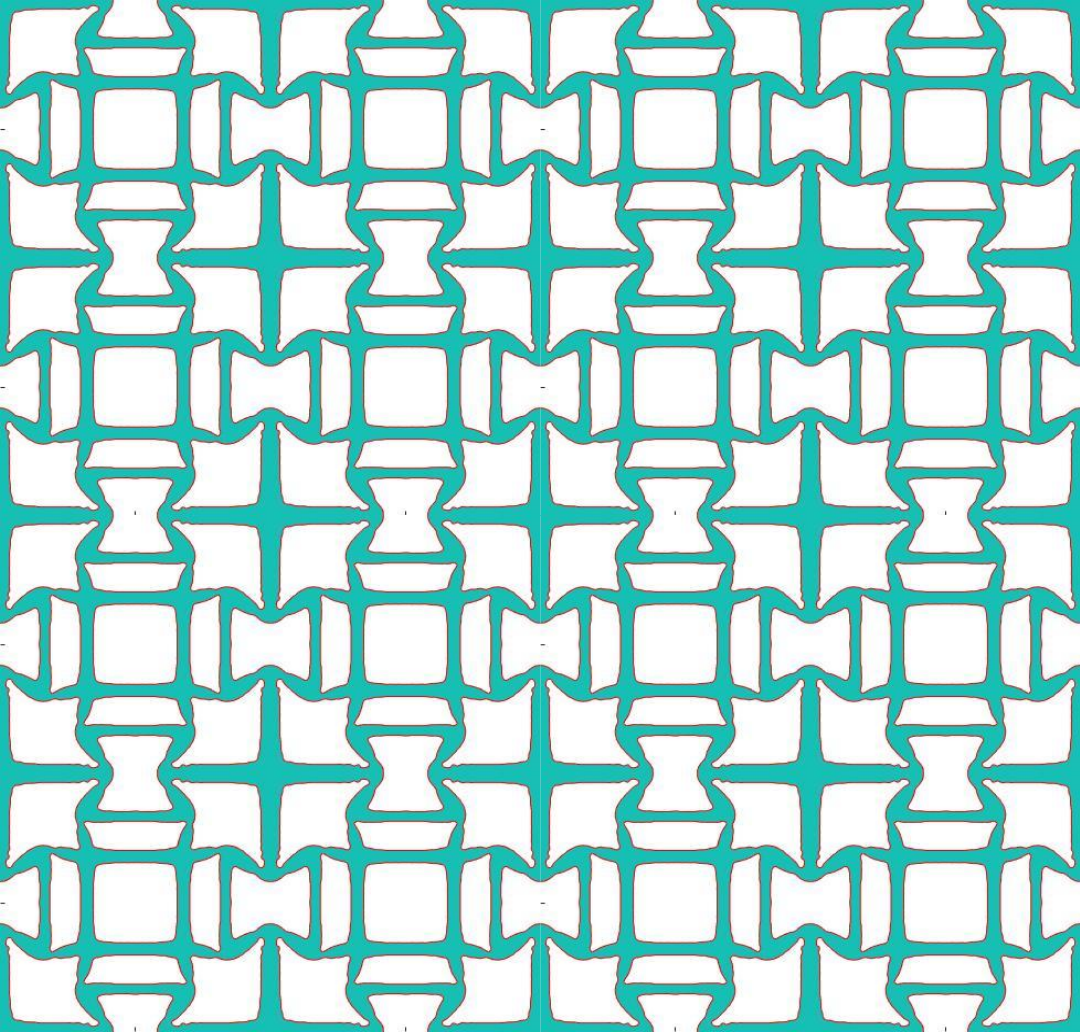}&  
    \scalebox{0.75}
    {
        $ 
        {\renewcommand{\arraystretch}{1.0}
            \begin{array}{*3{>{\displaystyle}l}} 
            \mathbf{C}^{H}=\left[\begin{array}{*3{>{\displaystyle}c}} 
            0.0840  & -0.0507 &0 \\
            -0.0507 & 0.0840 &0 \\
            0 & 0 &0.004 
            \end{array}\right]\\\\
            
            V_{f}=0.35\\\\
            \nu_{xy}=\nu_{yx}=-0.603
        \end{array}
    }
    $
} &\\
\hline\\

\end{tabular}
\captionof{figure}{Influence of initial configurations and volume fraction to final designs.}
\label{fig_optimized_designs_from_different_initial_configs}     
\end{center}

\begin{center}
\centering 
\setlength\figureheight{5.0cm}
\setlength\figurewidth{6.0cm}
\setlength\tabcolsep{0.0pt} 
\begin{tabular}{cc}
\includegraphics[width=0.825\linewidth]{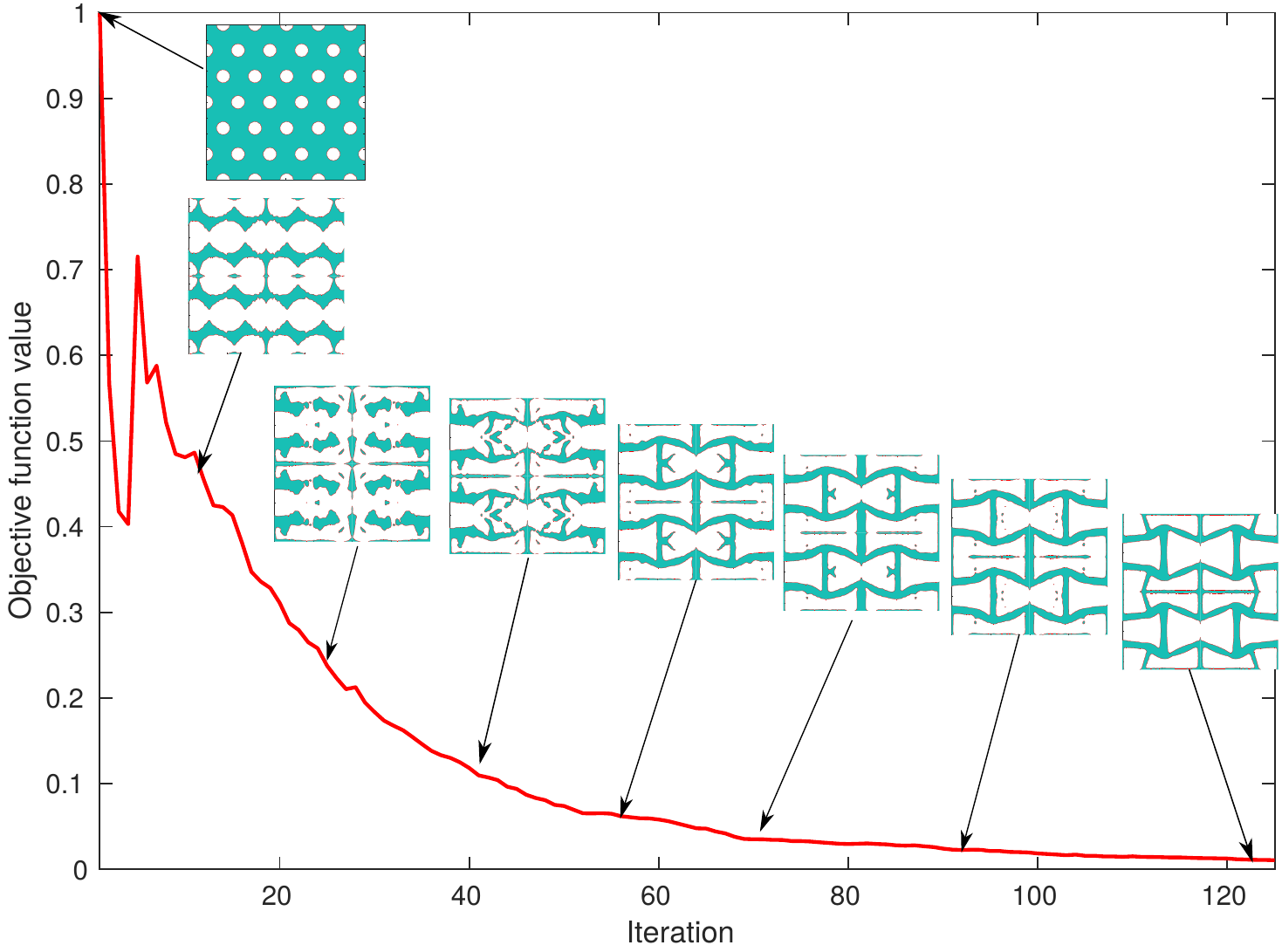}
\end{tabular}
\captionof{figure}{Convergence of the objective function.}
\label{fig_convergence_of_the_objective_function_without_rom}
\end{center}

\begin{center}
\centering 
\setlength\figureheight{5.0cm}
\setlength\figurewidth{6.0cm}
\setlength\tabcolsep{0.0pt} 
\begin{tabular}{cc}
\includegraphics[width=0.825\linewidth]{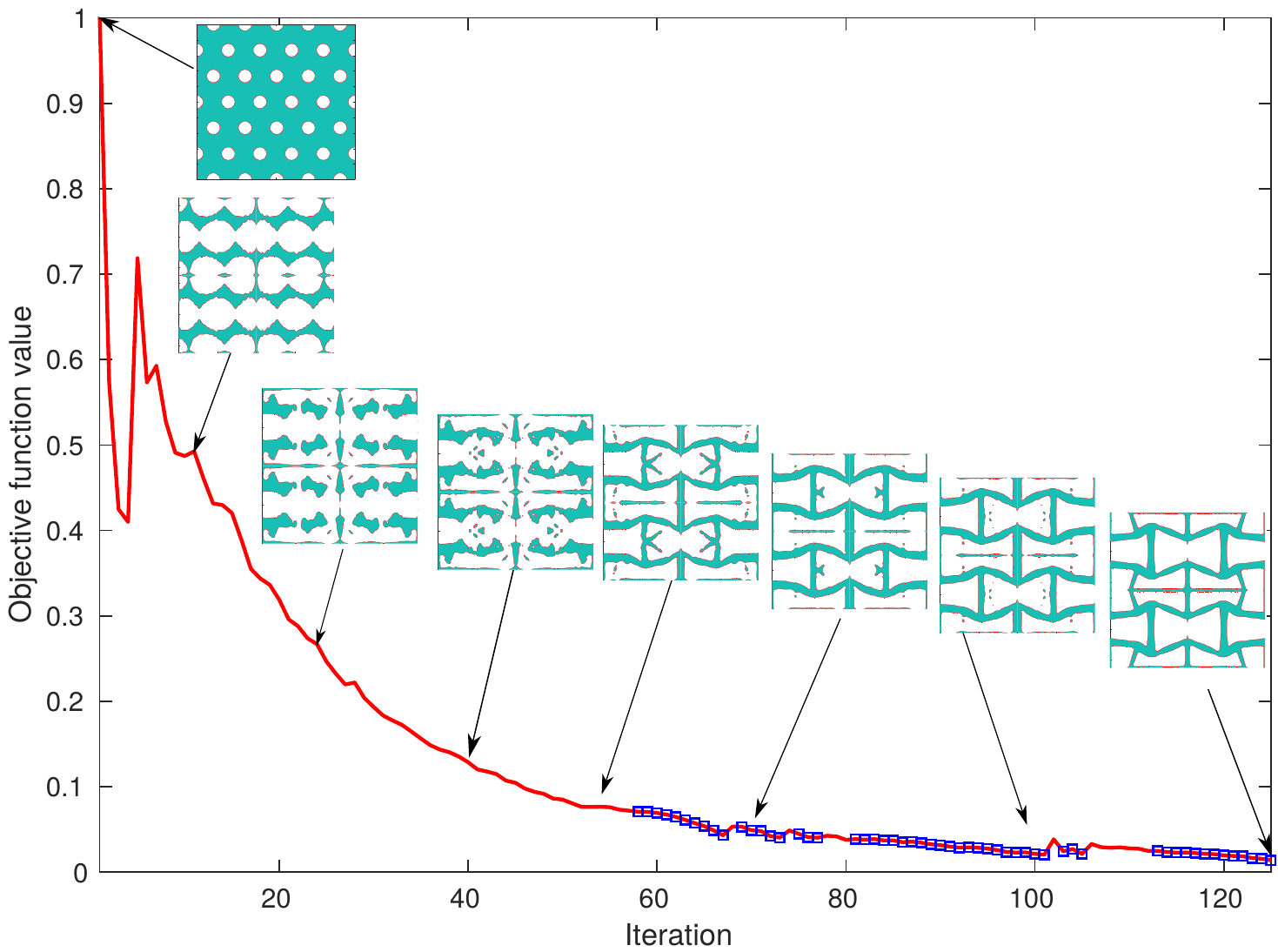}\\
\end{tabular}
\captionof{figure}{Convergence of the objective function by using the algorithm with reduced order model in \cref{fig_optimization_algorithm_with_reduced_basis} (right). Blue squares indicate the reduced solutions are used.}
\label{fig_convergence_of_the_objective_function_with_rom}
\end{center}

We further design three-dimensional metamaterials with negative Poisson's ratio by considering a cubic unit cell and investigate the efficiency of using reduced modeling when the size of problems is increased. Design domains are  discretized with $40^{3}$ quadratic B{\'e}zier elements  thus leading to moderately large models which are time consuming if balance equations are solved in  the finite dimensional space $V^{h}_{0}$.  We use a reduced basis with $nb=12$ and  errors smaller than $tol=0.01$ to decide whether the reduced solutions are accurate and subsequently employed to advance a new topology.  \Autoref{fig_optimized_designs_from_different_initial_configs_3d} provides optimized structures and corresponding homogenized elastic tensors which are results of optimization algorithm with reduced modeling. Designed geometries exhibit auxetic effect if the structures are under compression or tension. Due to  the nature  of gradient-based algorithms which  accept local optimums, final results can be different although the same target elastic tensor is set for the objective functions. 

The plot of convergence in \autoref{fig_convergence_of_the_objective_function_with_rom_3d} indicates iterations at which reduced modeling applied successfully. A similar behavior of the algorithm as shown in the two-dimensional problem  (cf. \autoref{fig_convergence_of_the_objective_function_with_rom}) is also observed in this 3D example. Significant changes topology and shapes lead to the reduced basis at the initial stage can not produce accurate solutions. Afterwards,  the shapes and topology advance gradually resulting on similar geometries produced and the reduced basis constructed from full solutions are able to reproduce accurate solutions used in next consecutive steps.

The computation time and number of iterations required in the 3D examples are summarized in \autoref{tab_CPU_times_compare_full_and-reduced_modeling_3d}. We  simulate  examples in  similar settings but with and without using the reduced order model. The optimized structures of each simulation are slightly different (and are not shown here) due to the local optimum characteristic of the optimizer. We purposely  eliminate all possible instabilities and keep the optimizer conservative by using small moving stepsizes \cite{Svanberg2007ab}, hence, the number of iterations is relatively large in all numerical examples. In this way, we can have adequate comparisons between the two algorithms and can emphasize the outperformance of the reduced order technique. The total CPU time required in full model is larger although fewer number of iterations are used. In such large scale problems, the time for stiffness matrix assembly  is negligible and the computational cost of solving the system of equation is dominant. If reduced solutions are accurate enough to advance new topology in the consecutive step, the computational cost of solving original large equations is eliminated by determining solutions of system with small size of $nb\times nb$  and a speed-up of factor 6.0 in solving the equation is achieved.

\begin{center}
\begin{tabular}{|c|c|c|c|c|c|}
\hline
\multirow{2}{*}{\begin{tabular}[c]{@{}c@{}}Unit cells\end{tabular}} & \multicolumn{2}{c|}{Full model} & \multicolumn{3}{c|}{Reduced order model} \\ \cline{2-6} 
& \begin{tabular}[c]{@{}c@{}}Number of\\ iterations\end{tabular} & \begin{tabular}[c]{@{}c@{}}CPU time\\ (in hour)\end{tabular} & \begin{tabular}[c]{@{}c@{}}Number of\\ iterations\end{tabular} & \begin{tabular}[c]{@{}c@{}}Number of iterations \\ using reduced solution\end{tabular} & \begin{tabular}[c]{@{}c@{}}CPU time\\ (in hour)\end{tabular} \\ \hline
a & 420 & 12.25 & 624 & 532 & 10.25 \\ \hline 
b & 672 & 19.3 & 685 & 628 & 10.9 \\ \hline  
c & 525 & 16.6 &  578 & 503 & 11.4 \\ \hline 
d & 480 & 15.2 & 472 & 403 & 8.9 \\ \hline 
\end{tabular}
\captionof{table}{Computation time of the 3D examples with and without using reduced order modelling.}
\label{tab_CPU_times_compare_full_and-reduced_modeling_3d}     
\end{center}

\begin{center}
\begin{tabular}{>{\centering\arraybackslash} m{3.0cm} >{\centering\arraybackslash} m{5.75cm} >{\centering\arraybackslash} m{0.8cm}  >{\centering\arraybackslash} m{3cm} >{\centering\arraybackslash} m{1.0 cm} }\hline
\small Initial design & \small Optimized unit cell  &  &\small Effective property $\mathbf{C}^{H}$ \& volume fraction  &\\
\hline\\

(a) \includegraphics[width=1.0\linewidth]{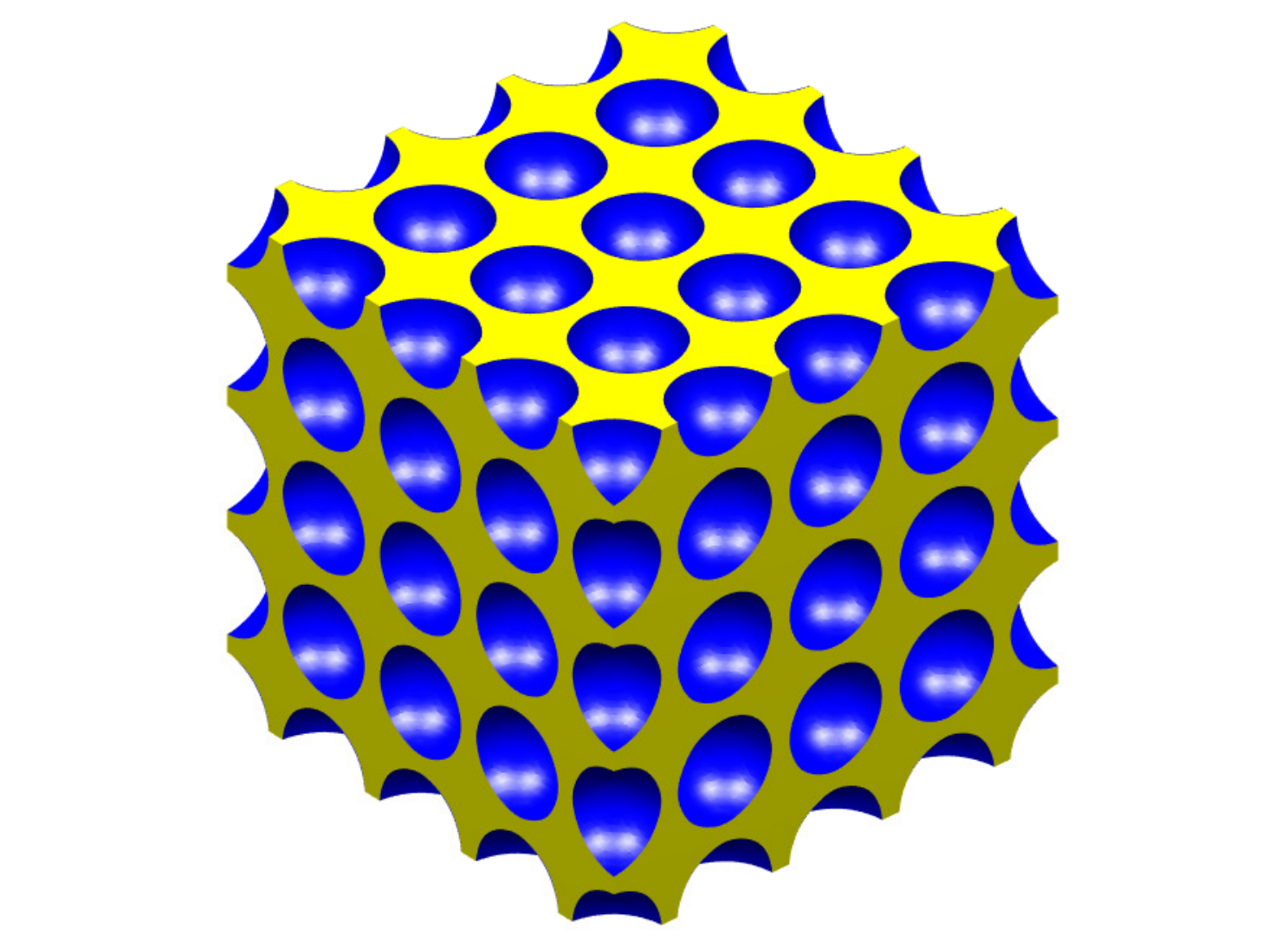} & 
\includegraphics[width=1.0\linewidth]{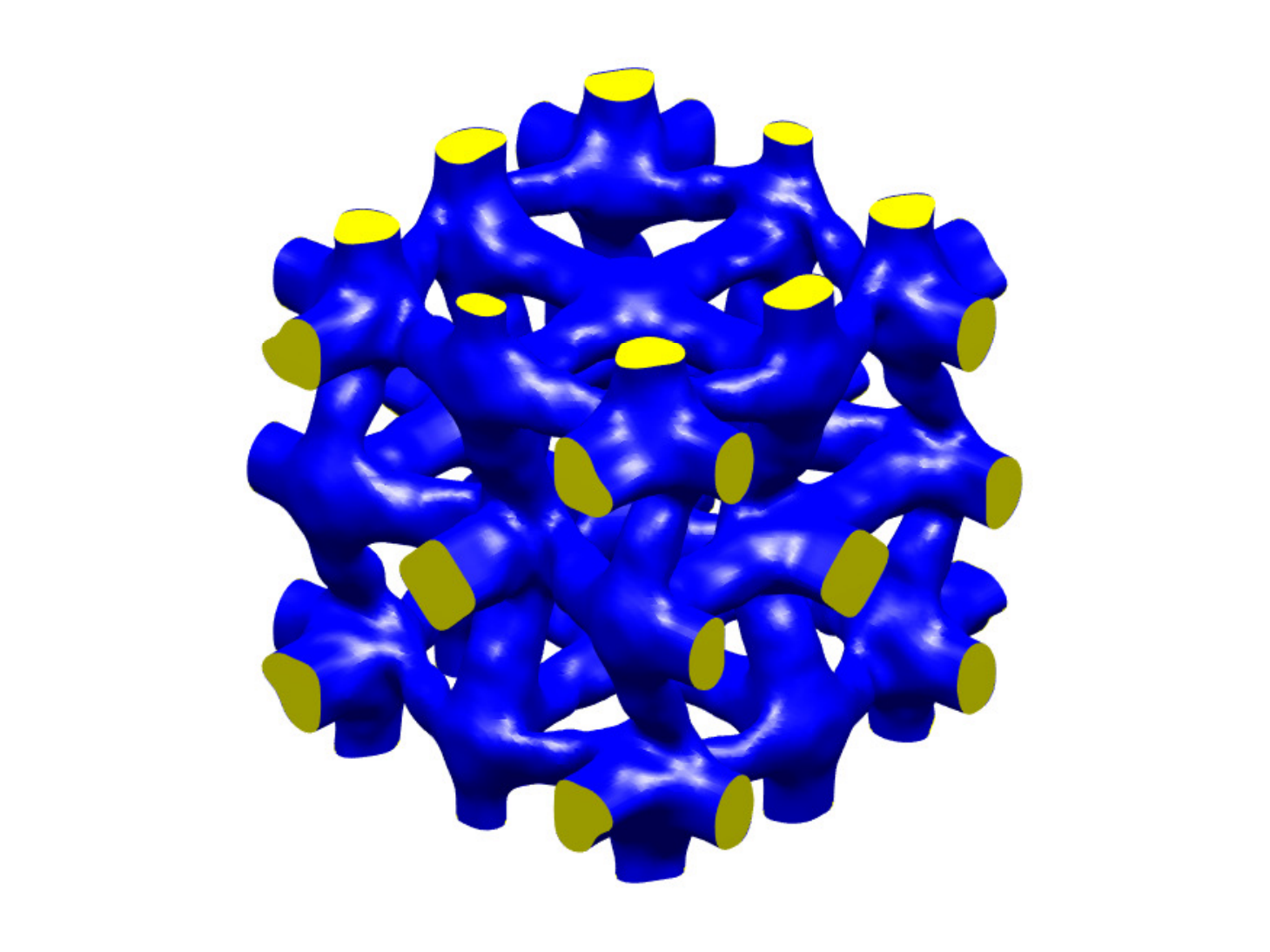}&  
\scalebox{0.6}
{
    $ 
    {\renewcommand{\arraystretch}{1.0}
        \begin{array}{*3{>{\displaystyle}l}} 
        0.1\times\left[\begin{array}{*6{>{\displaystyle}c}} 
        0.29 & -0.11 & -0.095 & 0 & 0 & 0 \\
        -0.11 & 0.38 & -0.063  & 0 & 0 & 0 \\
        -0.095 & -0.063 & 0.50  & 0  & 0 & 0 \\
        0 & 0 & 0 & 0.04  & 0 & 0\\
        0 & 0 & 0 & 0  & 0.03 & 0 \\
        0 & 0 & 0 & 0  & 0 & 0.03 
        \end{array}\right] \\\\
        V_{f}=0.225
        \end{array}
    }
    $
} &\\
\hline\\

(b) \includegraphics[width=1.0\linewidth]{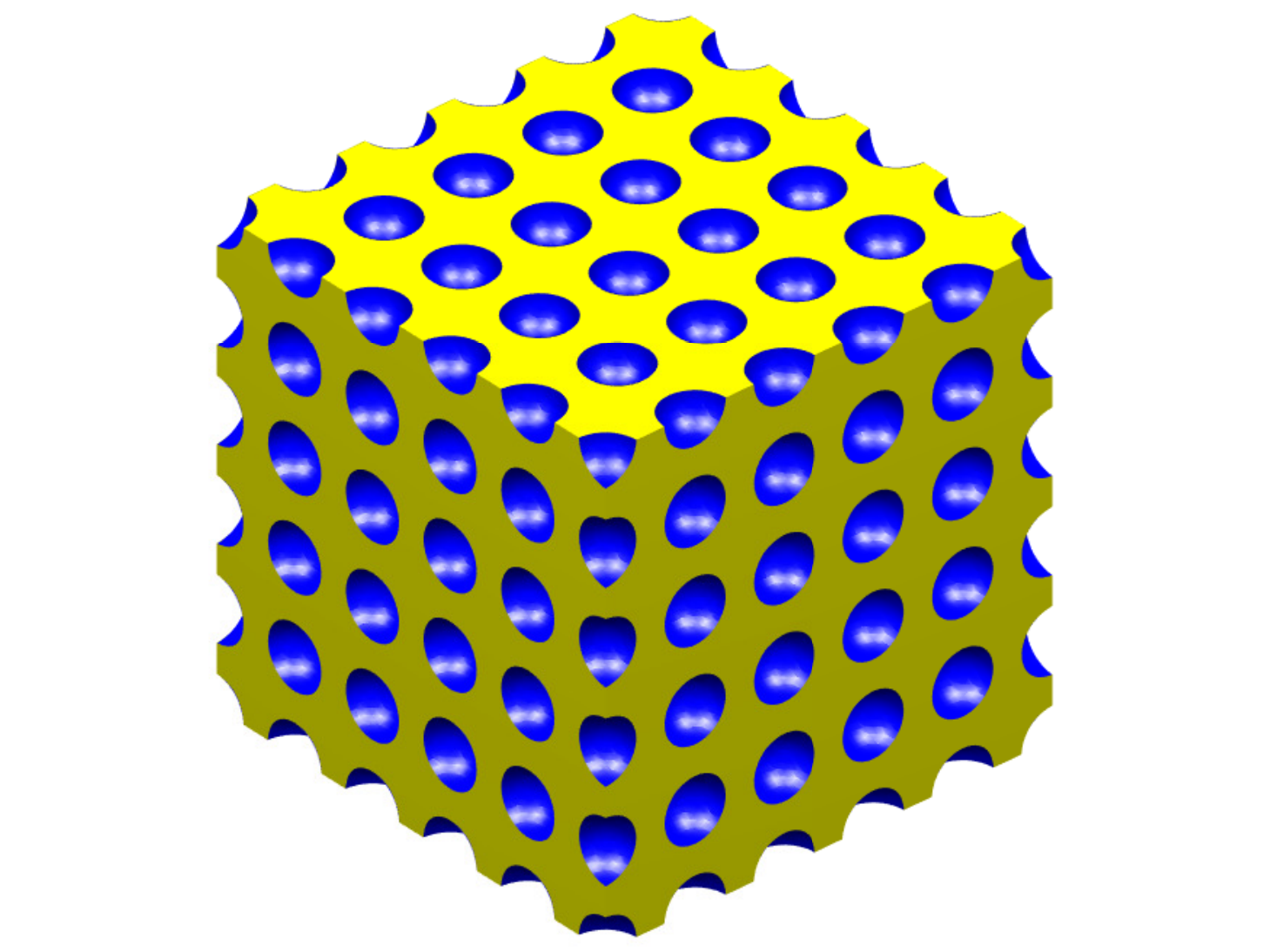} &
\includegraphics[width=0.9\linewidth]{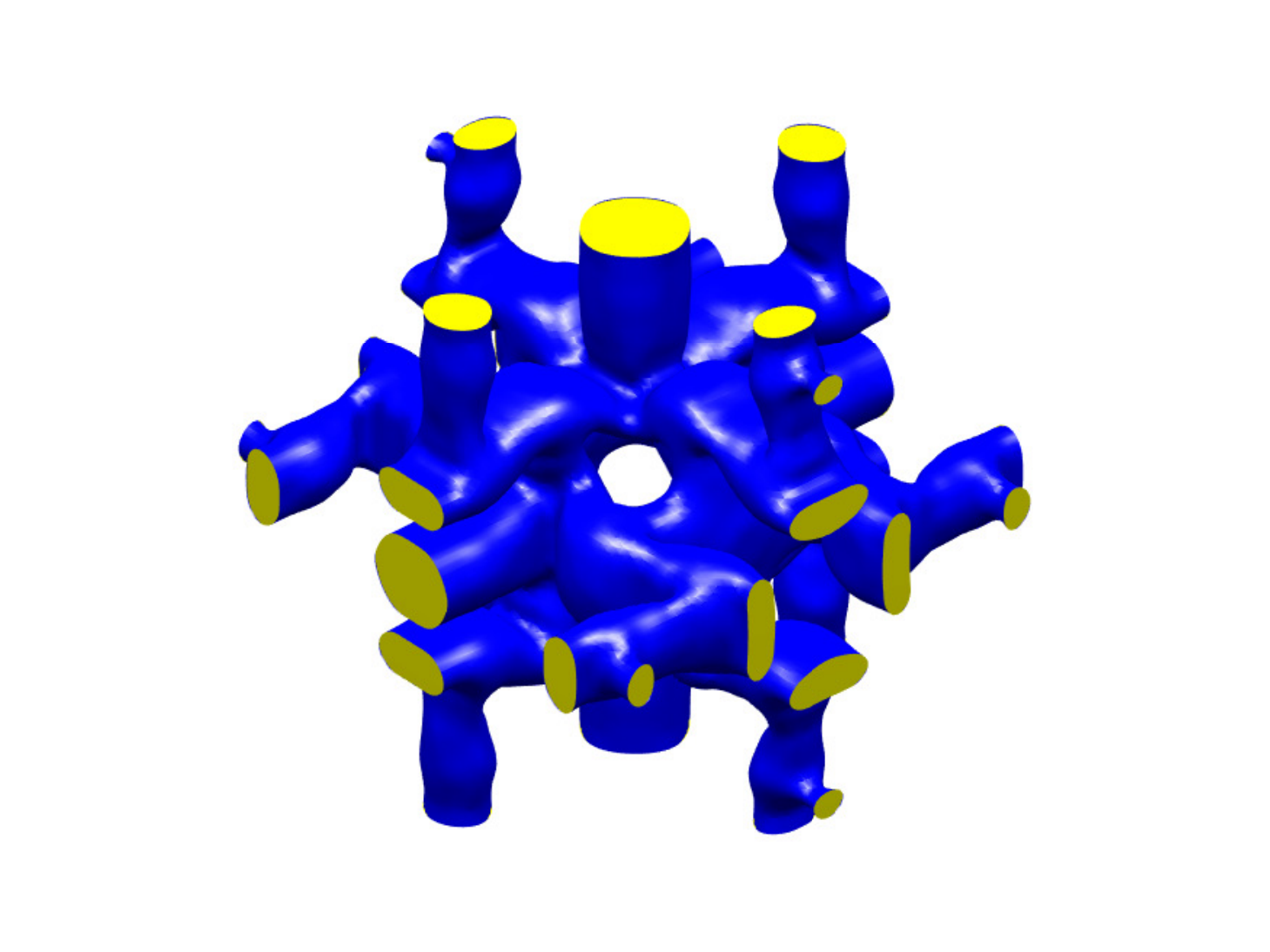}&  
\scalebox{0.6}
{
    $ 
    {\renewcommand{\arraystretch}{1.0}
        \begin{array}{*3{>{\displaystyle}l}} 
        0.1\times\left[\begin{array}{*6{>{\displaystyle}c}} 
        0.34 & -0.077 & -0.048 & 0 & 0 & 0 \\
        -0.077 & 0.4 & -0.053  & 0 & 0 & 0 \\
        -0.048 & -0.053 & 0.27  & 0  & 0 & 0 \\
        0 & 0 & 0 & 0.016  & 0 & 0\\
        0 & 0 & 0 & 0  & 0.013 & 0 \\
        0 & 0 & 0 & 0  & 0 & 0.017 
        \end{array}\right] \\\\
        V_{f}=0.25
        \end{array}
    }
    $
} &\\
\hline\\


(c) \includegraphics[width=1.0\linewidth]{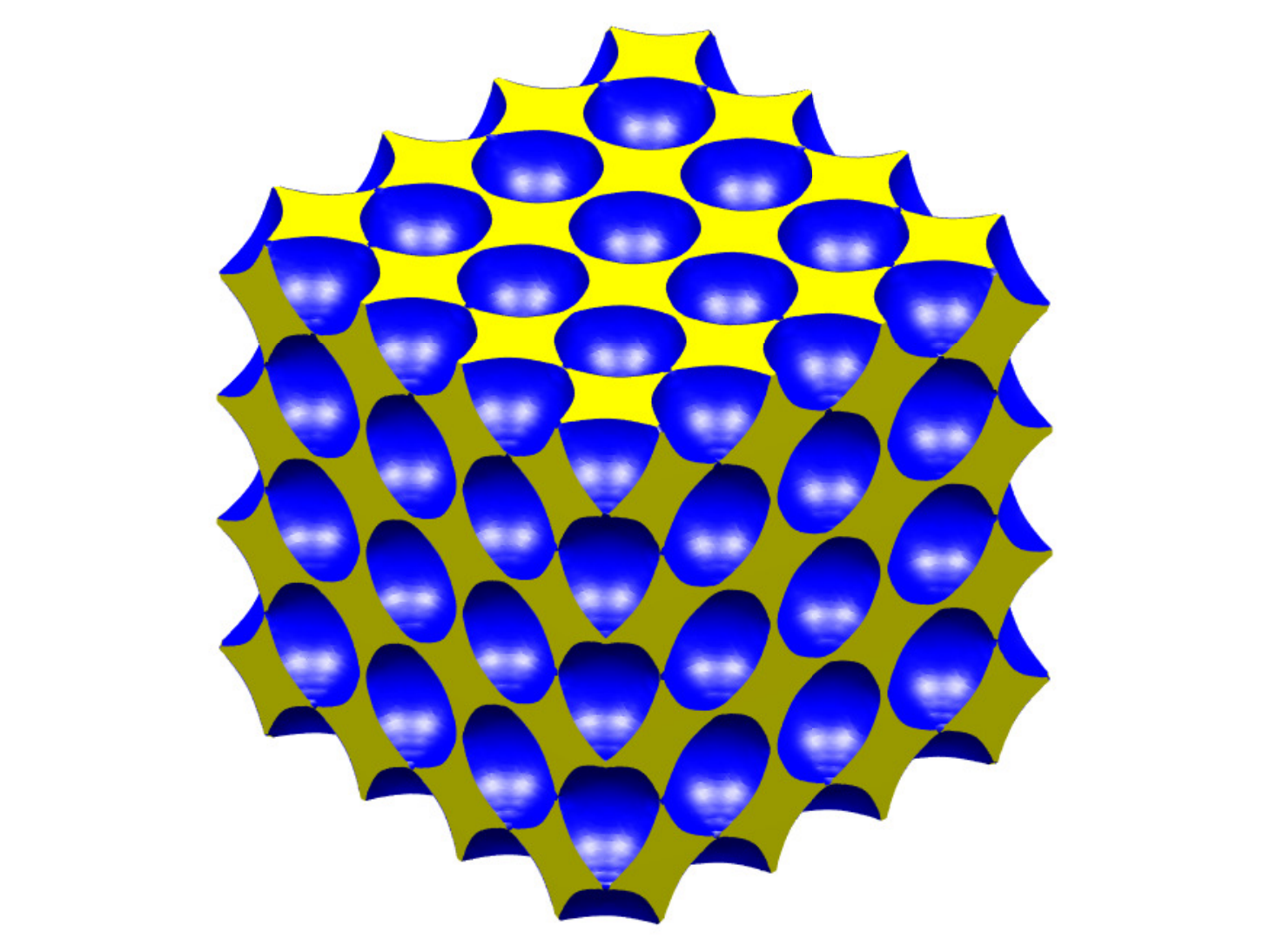} & 
\includegraphics[width=1.0\linewidth]{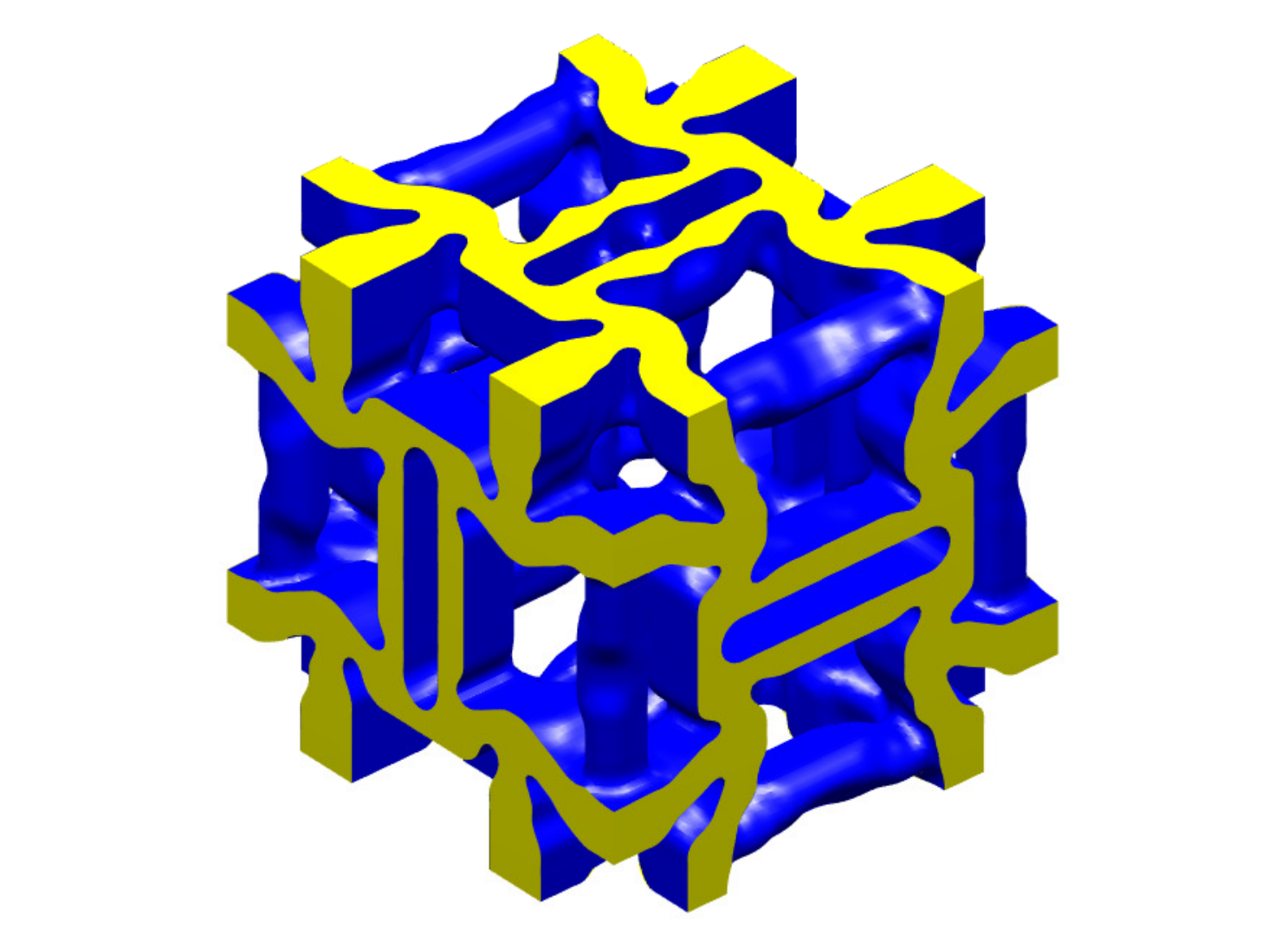}&  
\scalebox{0.6}
{
    $ 
    {\renewcommand{\arraystretch}{1.0}
        \begin{array}{*3{>{\displaystyle}l}} 
        0.1\times\left[\begin{array}{*6{>{\displaystyle}c}} 
        0.5 & -0.086 & -0.083 & 0 & 0 & 0 \\
        -0.086 & 0.65 & -0.09  & 0 & 0 & 0 \\
        -0.083 & -0.09 & 0.61  & 0  & 0 & 0 \\
        0 & 0 & 0 & 0.046  & 0 & 0\\
        0 & 0 & 0 & 0  & 0.047 & 0 \\
        0 & 0 & 0 & 0  & 0 & 0.052 
        \end{array}\right] \\\\
        V_{f}=0.3
        \end{array}
    }
    $
} &\\
\hline\\

(d) \includegraphics[width=1.0\linewidth]{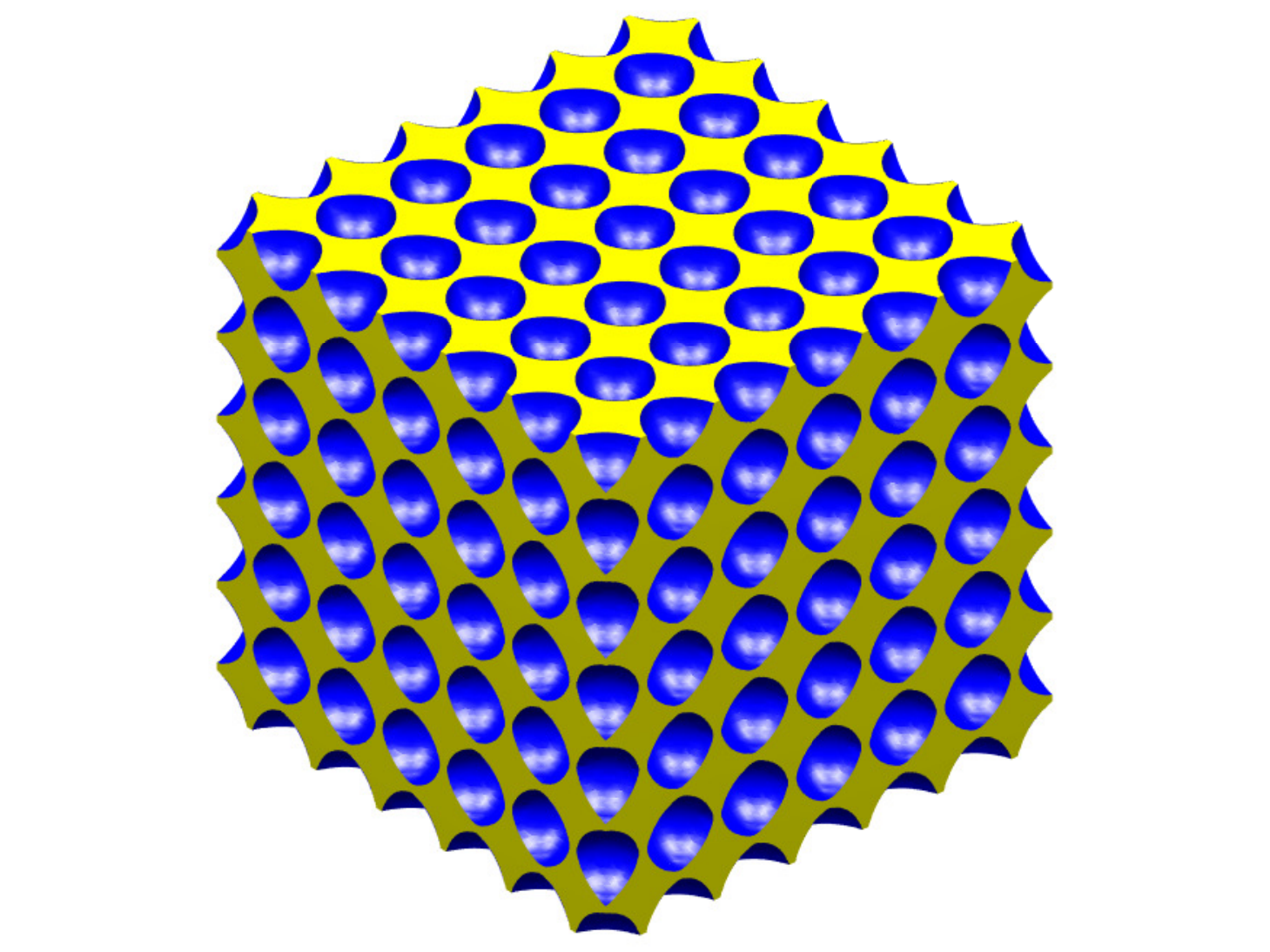} & 
\includegraphics[width=1.0\linewidth]{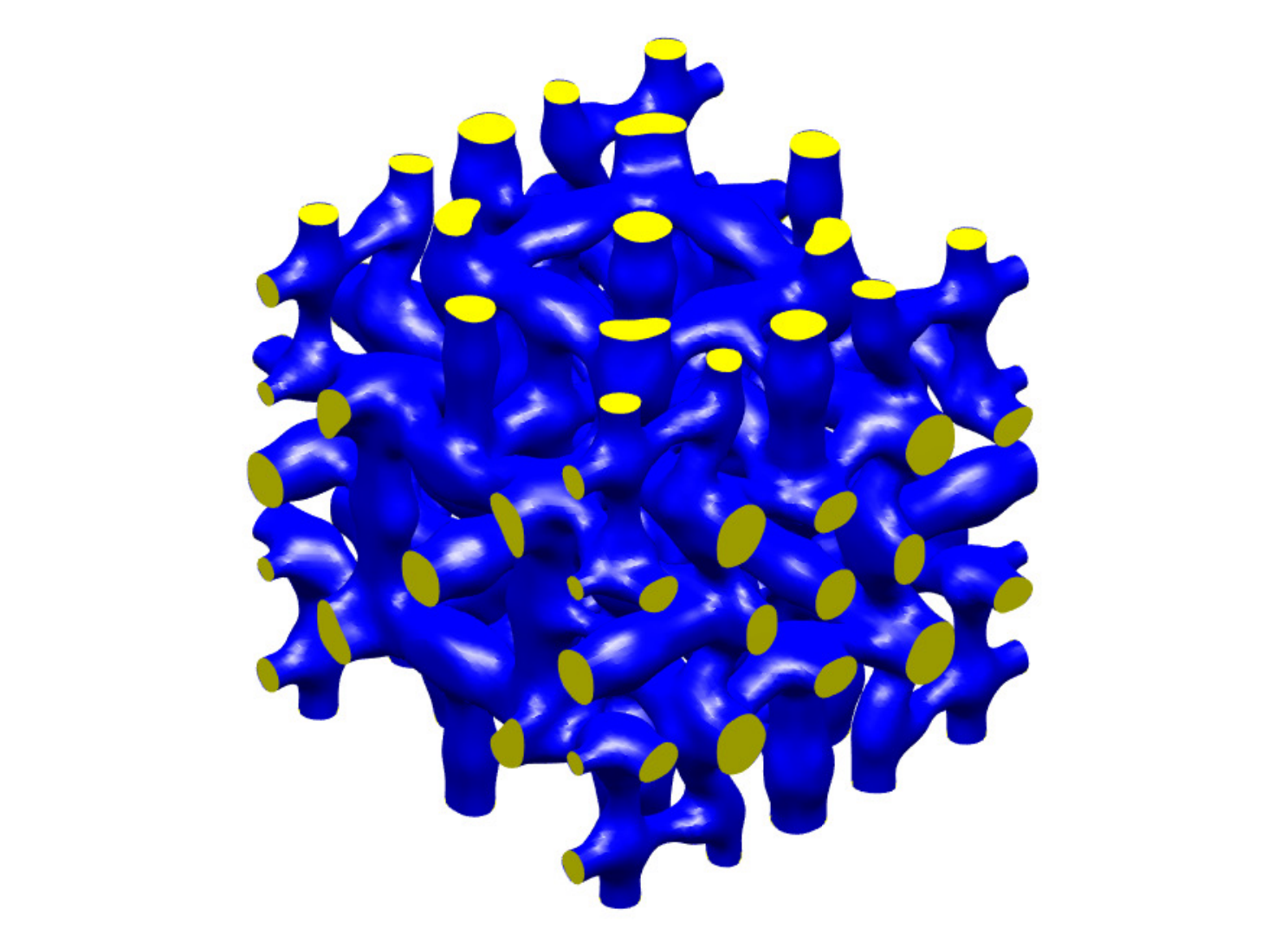} &  
\scalebox{0.6}
{
    $ 
    {\renewcommand{\arraystretch}{1.0}
        \begin{array}{*3{>{\displaystyle}l}} 
        0.1\times\left[\begin{array}{*6{>{\displaystyle}c}} 
        0.44 & -0.069 & -0.092 & 0 & 0 & 0 \\
        -0.069 & 0.51 & -0.012  & 0 & 0 & 0 \\
        -0.092 & -0.012 & 0.61  & 0  & 0 & 0 \\
        0 & 0 & 0 & 0.038  & 0 & 0\\
        0 & 0 & 0 & 0  & 0.042 & 0 \\
        0 & 0 & 0 & 0  & 0 & 0.05 
        \end{array}\right] \\\\
        V_{f}=0.3
        \end{array}
    }
    $
} &\\
\hline\\
\end{tabular}
\captionof{figure}{Influence of initial configurations and volume fractions to final designs in three-dimensional metamaterials design.}
\label{fig_optimized_designs_from_different_initial_configs_3d}     
\end{center}

\begin{center}
\centering 
\setlength\figureheight{5.0cm}
\setlength\figurewidth{6.0cm}
\setlength\tabcolsep{0.0pt} 
\begin{tabular}{cc}
\includegraphics[width=1.0\linewidth]{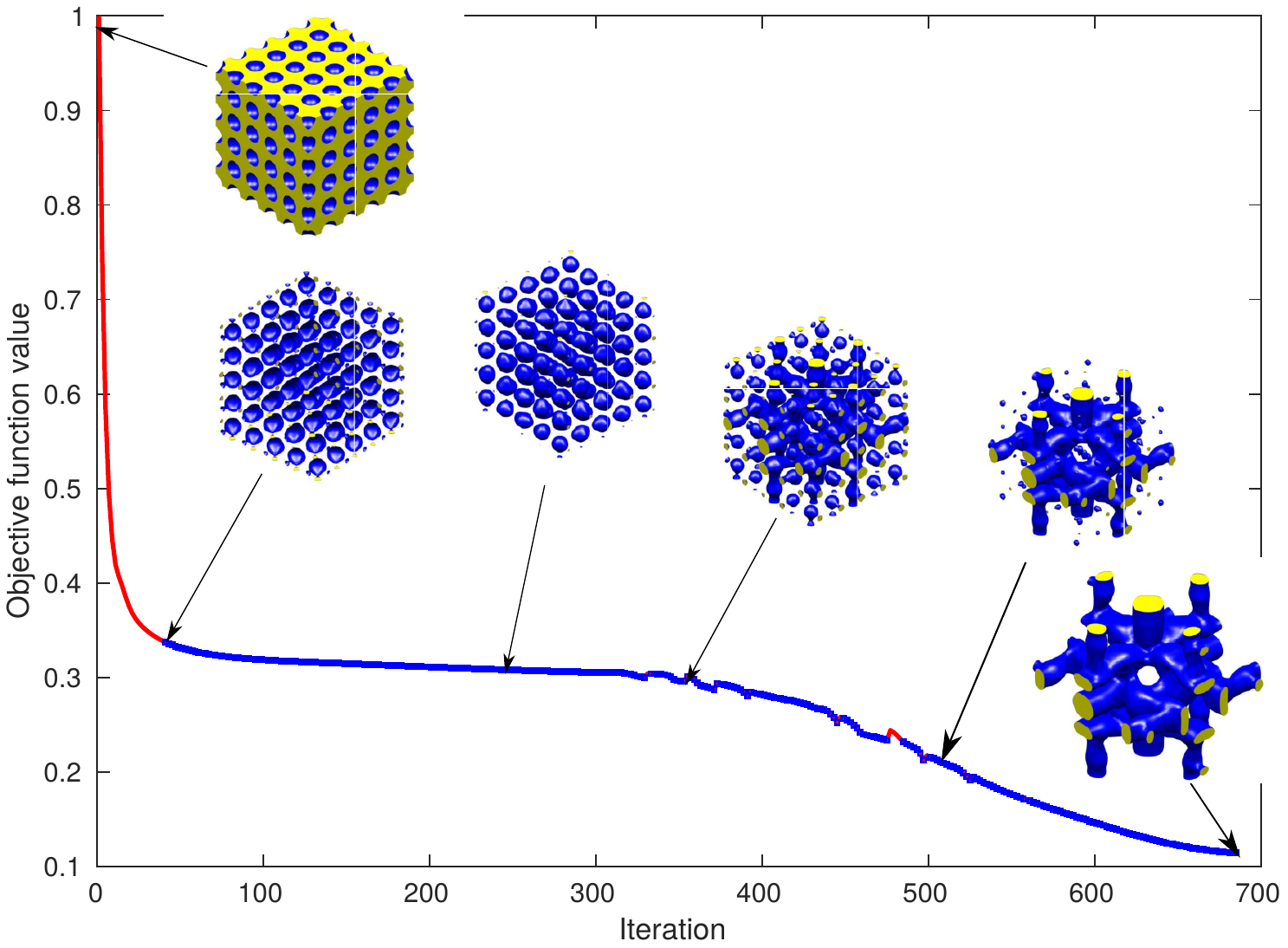}
\end{tabular}
\captionof{figure}{Convergence of the objective function (of the unit cell (b) in \autoref{fig_optimized_designs_from_different_initial_configs_3d}) by using the algorithm with reduced order modeling. Blue squares indicate the reduced solutions are used.}
\label{fig_convergence_of_the_objective_function_with_rom_3d}
\end{center}

The feasibility of using reduced solutions in shape and topology optimization  can trace back to reduced order models using the offline strategy. In these approaches, a set of parameters can represent the geometry adjustment, and  pre-calculated solutions of the model are obtained by evaluating at selected parameter values. It is noted that when the variation of parameters does not result in the enormous change of geometry shape or topology, this allows the reduced basis to generate accurate solutions. In this current work, a similar idea for constructing the reduced basis is employed. Indeed, the stored solutions obtained from geometries are used to construct the reduced subspace. The amount of difference in geometries decides the enrichment level of the reduced basis and the accuracy of reduced solutions. The method is practical since minor changes in geometries always appear after several iterations (see \autoref{fig_convergence_of_the_objective_function_with_rom} and \autoref{fig_convergence_of_the_objective_function_with_rom_3d}).

\section{Conclusion}\label{sect_conclusion}
Material microstructures with negative effective Poisson's ratio have been designed by topology optimization. Asymptotic homogenization method and sensitivity analysis of the effective elasticity tensor was revised and numerically implemented by isogeometric analysis with B{\'e}zier extraction. Geometries of the unit cell were represented by a parameterized level set function which allows flexible changes in shape and topology designs, and optimized structures having clear boundaries in order to be able to make prototypes with 3D printers directly. A reduced order modeling combined with topology optimization was introduced to improve the computational efficiency in solving the linear system of equations. The numerical results show that  such a combination are beneficial in  topology optimization and could be extended to other physical problems such as improving electrical coefficients, or negative thermal expansion materials. 

%

        

    
    
    \section*{References}

    
    \bibliography{library}
    
    \bibliographystyle{unsrt}

    
    
\end{document}